\documentclass[aps,pre,preprint,groupedaddress,amsmath,amssymb,floatfix]{revtex4-1}

\usepackage{graphicx}

\begin{document}

\title{Formation of Anti-Waves in Gap Junction Coupled Chains of Neurons}
\author{Alexander Urban
 }
\affiliation{Department of Physics, University of Pittsburgh,100 Allen Hall
3941 O'Hara Street, 
Pittsburgh, PA 15260}
\author{Bard Ermentrout}
\affiliation{Department of Mathematics, University of Pittsburgh, 139 University Place, Pittsburgh, PA 15260}

\date{\today}

\begin{abstract}
Using network models consisting of gap junction coupled Wang-Buszaki neurons, we demonstrate that it is possible to obtain not only synchronous activity between neurons but also a variety of constant phase shifts between $0$ and $\pi$. We call these phase shifts {\it{intermediate stable phase-locked states}}. These phase shifts can produce a large variety of wave-like activity patterns in one-dimensional chains and two-dimensional arrays of neurons, which can be studied by reducing the system of equations to a phase model. The $2\pi$ periodic coupling functions of these models are characterized by prominent higher order terms in their Fourier expansion, which can be varied by changing model parameters. We study how the relative contribution of the odd and even terms affect what solutions are possible, the basin of attraction of those solutions and their stability. These models may be applicable to the spinal central pattern generators of the dogfish and also to the developing neocortex of the neonatal rat.
\end{abstract}
\maketitle
\section{Introduction}

In the past few decades, neuroscientists have discovered the importance of oscillations in the brain. 
Oscillations occur in many neural networks and play key roles in sensory as well as motor systems \cite{grillner}\cite{izhikevich}. On the smallest scale of the nervous system, individual neurons can behave as complex nonlinear oscillators.\cite{buszaki} On a larger scale,  oscillatory patterns in EEG (electroencephalographic) recordings have been shown to correspond to different cognitive states\cite{buszaki}. Phase oscillator models (obtained from systems of weakly coupled oscillators) have proven extremely useful in the analysis of such systems \cite{cohen}\cite{ermentrout_gelperin}.
 
A pair of symmetric weakly coupled oscillators always has at least two possible periodic (phase-locked) states: the synchronous state and the anti-phase atate (where they are a half a cycle out of phase). 
 Suppose that for some range of parameters the synchronous solution is stable and as one of these parameters is varied, the synchronous solution loses stability. In a symmetrically coupled system the synchronous solution generically loses stability in a pitchfork bifurcation. When this pitchfork bifurcation is supercritical it gives rise to two  new stable phase-locked solutions in which the phase difference between oscillators is some constant between $0$ and $\pi$. We refer to such solutions as  {\it{Intermediate stable phase locked states}}. Intermediate stable phase locked states between pairs of neurons can be produced in a variety of neuron models by using a variety of coupling schemes as well as adjusting the parameters which affect the excitability of individual neurons \cite{adaptation}\cite{Mancilla}\cite{Vreeswijk}. The first work examining this behavior in conductance-based models was by Vanvreeswijk et al \cite{Vreeswijk}.  In this paper, the authors studied integrate-and-fire as well as Hodgkin-Huxley neurons. Cymbalyuk has both modeled and experimentally demonstrated an intermediate stable phase-locked state in a system of two ``silicon neurons"\footnote{A silicon neuron refers to an electronic circuit designed to mimic a biological neuron}\cite{cymbalyuk}. These intermediate phase-locked states can lead to wave-like behavior in large networks.

Wave activity is ubiquitous in the brain. Traveling waves of electrical activity in the brains of animals have been observed in a variety of species and occur in a diverse set of structures. These structures include the retina, olfactory cortex \cite{limax}, peri-geniculate nucleus\cite{kim}, neocortex \cite{peinado} and spinal cord \cite{grillner}, among others \cite{kleinfeld},\cite{kim},\cite{peinado},\cite{limax},\cite{ratspinalcordwaves}\cite{kim}. In phase models, waves correspond to constant nonzero phase-differences between successive pairs of oscillators. There are many ways to generate such phase differences (see \cite{bardbook}, section 8.3.5) such as a gradient in natural frequencies \cite{cohen}, pacemakers, or manipulation of the boundary conditions \cite{chains}. In some swimming organisms, it is very important that a fixed phase-lag be maintained over a wide range of frequencies. For example, in the lamprey, the lag is about $2\pi/100$ \cite{cohen1992}, while for the crayfish it is $2\pi/4$ \cite{jones2003}. As we will show in this paper, intermediate stable phase-locked states provide a simple way to produce waves with a stable fixed inter-oscillator phase-difference.

The paper is organized as follows. We first show that by varying the excitability of gap-junction coupled Wang-Buszaki neurons, we can produce intermediate stable phase-locked states.  We then explore the consequences of these states in one-dimensional nearest-neighbor coupled chains. We prove the existence and stability of a wide variety of complex waves including simple waves and zig-zag waves. We then study two-dimensional arrays and find two-dimensional analogues of zig-zag and regular waves.

\section{Gap Junction Coupling between pairs of Wang-Buszaki Neurons}
\subsection{Measuring the Phase Difference Between Spiking Neurons}\label{model}
The Wang-Buszaki model is a conductance based neuron model derived from the Hodgkin Huxley model. It was originally used to describe fast spiking interneurons in the hippocampus \cite{wang}. In this section we examine pairs of Wang-Buszaki neurons reciprocally coupled with gap junctions and ask what happens as we vary parameters which affect the excitability of the individual cells.
Gap junctions are specialized ion channels connecting the cytoplasm of the pre and post-synaptic cell. A depolarizing ionic current is driven by the potential difference between the cells. 
In order to understand the role that gap junctions play in rhythmically oscillating networks, consider just two neurons coupled with gap junctions. 
The coupled neuron equations will have the following form:
\begin{eqnarray}
C\dot{V_{1}}&=&-I_{Na}-I_k-I_L-I_0+g(V_2-V_1)\nonumber\\
C\dot{V_{2}}&=&-I_{Na}-I_k-I_L-I_0+g(V_1-V_2)\nonumber\\
\end{eqnarray}
These currents are given by the equations:
\begin{eqnarray}
I_{Na}&=&g_{Na}m^3_{\infty}h(V-V_{Na})\nonumber\\
I_{k}&=&g_{k}n^4(V-V_k)\nonumber\\
I_L&=&g_L(V-V_L)
\end{eqnarray}
C is the the membrane capacitance measured in units of $\mu F/cm^2$. The parameters: $V_{Na}$,$V_k$ and $V_L$ are the reversal potentials for the ion channels. The parameters $g,g_{Na},g_k$ and $g_L$ are constants describing the conductances of their respectively labeled ion channels. Typically, they are measured in units of $mS/cm^2$. The $m$ and the $h$ are the time-dependent activation and decativation variables which are described by the equations:
\begin{eqnarray}
\frac{dh}{dt}&=&\eta(\alpha_h(V)(1-h)-\beta_h(V)h)\nonumber\\
\frac{dm}{dt}&=&\eta(\alpha_m(V)(1-m)-\beta_m(V)m)
\end{eqnarray}
In these equations, $\eta$ is an overall channel switching rate determined by the temperature of the system\footnote{
\begin{math}\eta=Q_{10}^{T-T_{base}}\end{math}: in this equations, $Q_{0}$ is the ``ratio of the rates for an increase in temperature of $10^0$C'' \cite{bardbook}. }. The nonlinearities $\alpha_i(V),\beta_i(V)$, as well as  the parameters used in the simulations are referenced in the Appendix \ref{model_eqns}. 

\begin{figure}
\centering
\includegraphics[width=6.5in, height=4.5in]{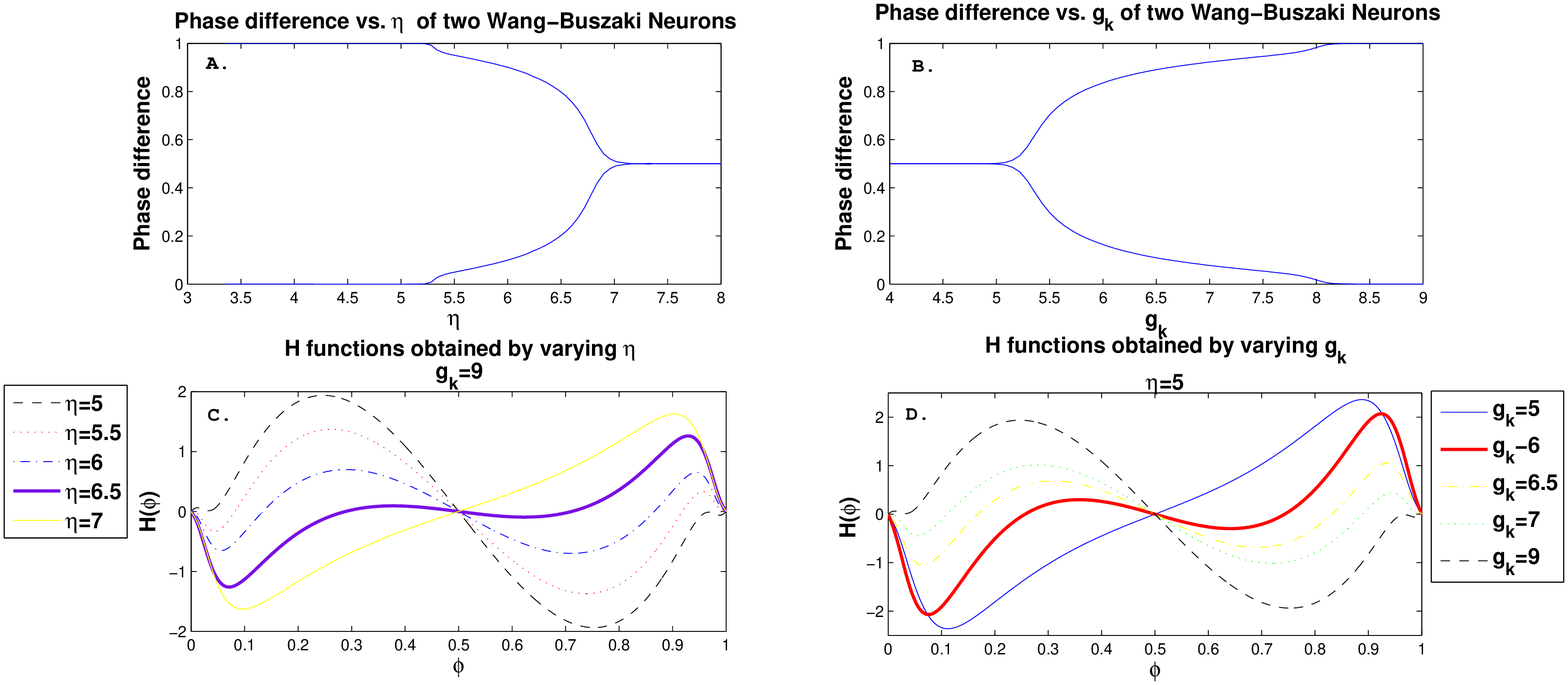}
\caption{ {\bf{A}}. The phase difference between two Wang Buszaki neurons (Full model) as a function of the parameter $\eta$. The diagram clearly illustrates a pitchfork bifurcation connecting the synchronous and anti-phase solutions. {\bf{B.}} Plot of the phase difference for varying values of $g_k$.  {\bf{C.}} Plot of the odd part of the  interaction function (cf equations (\ref{eq:hfun0},\ref{eq:pdiff})) for different values of $\eta$ (the dimensionless temperature dependent time constant). {\bf{D.}} Calculation of the odd part of the interaction function for different values of $g_{k}$. The zeros of these functions correspond to the stable phase locked solutions.}\label{fig:fig3}
\end{figure}

Previously, Pfeuty et al. have studied pairs of Wang-Buszaki neurons with gap junction coupling and shown that by varying the potassium and sodium conductances, one can vary the stability of the synchronous in-phase and anti-phase states \cite{Pfeuty}. In order to find parameter regimes in which intermediate phase-locked states exist, we varied the potassium conductance, $g_k$  and also the temperature dependent rate constant, $\eta.$  Parameters such as these play key roles in the behavior of both central pattern generators and large scale oscillatory networks \cite{Nusbaum}\cite{javedan}. The most important consequence of varying these parameters is that the absolute refractory period decreases. The neuron is less hyperpolarized after the action potential for larger $\eta$ or smaller $g_K$ \cite{Pfeuty}. In other words, this smaller absolute refractory region allows for the neuron to fire more quickly \cite{Pfeuty}. Consider the upper half of Figure \ref{fig:fig3}: Panels A and B are calculations of the phase difference between two Wang-Buszaki neurons computed after $1500$ ms of integration time for different values of $g_{k}$ (the maximal potassium conductance) and $\eta$ (the temperature dependent time constant). The phase difference, $\phi$  was measured as  the difference between the  times at which $V_{1}$ and $V_{2}$ cross zero with a positive slope divided by the period of the oscillation. Holding the stimulus current constant at $i_{0}=0.63$nA and varying either the parameter $\eta$ or $g_{k}$, we are able to demonstrate a supercritical pitchfork bifurcation in the phase difference between neurons. Figure \ref{fig:fig3}A  illustrates the bifurcation of the system of two neurons from synchrony to anti-phase behavior as $\eta$ is increased. For values of $\eta$ between $5.0$ and $7.0$ the system passes through all possible stable relative phase-locked solutions varying between $0.0$ and $\pi$. Figure \ref{fig:fig3}B  is a plot of the phase difference between neurons as we decrease $g_{k}$. As we decrease $g_{k}$ between $8.0$ and $5.0$ we also obtain intermediate stable phase-locked solutions.

\subsection{Calculation Of The Interaction Function}
In order to simplify the analysis, we reduce our system of neurons to a phase model.
This can be accomplished by applying Malkins Theorem \cite{izhikevich}.
 Consider two weakly coupled systems (By weak, we mean that the coupling acts to only affect the phase and not the amplitude of the oscillation \cite{izhikevich}) of coupled differential equations describing tonically firing neurons:
\begin{eqnarray}\label{voltage}
V_{1}' &=& F(V_1) +\epsilon C_1(V_1,V_2) + O(\epsilon^2) \\
V_{2}' &=& F(V_2) +\epsilon C_2(V_{2},V_{1}) + O(\epsilon^2).
\end{eqnarray}
Then Malkin's theorem states:\\ 
$V_{1}(t)=V_0(\theta_1)+O(\epsilon)$,
$V_{2}(t)=V_0(\theta_2)+O(\epsilon)$\\
where $V_0$ is the solution to the homogenous equation. Furthermore:
\begin{eqnarray}
\frac{d\theta_{1}}{dt} &=& 1 + \epsilon H_1(\theta_2-\theta_1) +
O(\epsilon^2)  \\
\frac{d\theta_{2}}{dt} &=& 1 + \epsilon H_2(\theta_1-\theta_2)  +
O(\epsilon^2) 
\end{eqnarray}
where:
\begin{equation}
\label{eq:hfun0}
H_j(\phi):=\frac{1}{T} \int_0^P V^*(t)\cdot C_j[V_0(t),V_0(t+\phi)]\ dt.
\end{equation}
Here $V^*$ is the adjoint of the linearized equation (\ref{voltage}).
Let $\phi:=\theta_2-\theta_1.$ Then:
\begin{equation}
\label{eq:pdiff}
\frac{d\phi}{dt} = -2 \epsilon H_{odd}(\phi):=\epsilon[H(-\phi)-H(\phi)].
\end{equation}
The right hand sides of these phase model equations are described by interaction functions which are derived from the full model. Zeros of $H_{odd}(\phi)$ determine the possible phase-locked states and those for which $H_{odd}'(\phi)>0$ are stable. 

We calculated the interaction functions for several values of $g_k$ and $\eta$. Figure 1C and 1D shows the odd portions of the interaction functions calculated with various parameter values.   
We write these interaction functions in terms of their Fourier series:
\begin{eqnarray}
H(x)=\frac{a_{0}}{2}+\sum^{\infty}_{n=1}\bigg( a_{n}\cos(nx)+b_{n}\sin(nx) \bigg)
\end{eqnarray}
For example, Table \ref{fourier} shows the first few Fourier modes of the interaction function computed from the full model with $\eta=6$. This table demonstrates the substantial contribution of higher order Fourier terms to the interaction function near the bifurcation point.

\begin{table}[ht]
\centering
\begin{tabular}{c c}
\hline\hline
$a_0=5.1974931$&\hspace{2.5in}\\
$a_1=-2.9970722$&$b_1=0.47408548$\\
$a_2=-0.92187762$&$b_2=-0.36833799$\\
$a_3=-0.44113794$&$b_3=-0.2577318$\\
$a_4=-0.25482759$&$b_4=-0.15762125$ \\
$a_5=-0.16416954$&$b_5=-0.09083201$\\
$a_6=-0.11295291$ & $b_6=-0.048487604$\\[1ex]
\hline
\end{tabular}
\caption{\label{fourier}Fourier Coefficients of the Interaction function, equation~(\ref{eq:hfun0})  (computed from the full model with $\eta=6$). Near the bifurcation point from synchrony there are substantial higher order even and odd Fourier terms. }
\end{table}
Observe that the first two odd Fourier terms in Table \ref{fourier} are the dominant ones ($b_1,b_2$). As the shape of $H_{odd}(\phi)$ depends only on the odd terms, $a_j$ have no effect on the existence and stability of the locked solutions for a pair of symmetrically coupled oscillators. However, once there are more than just two oscillators, the even terms play an important role, particularly as fasr as stability is concerned. We find that as $\eta$ changes between 5 and 7, the Fourier coefficient, $b_2$ remains negative and $b_1$ goes to zero. For this reason, in the rest of this paper, we will use a truncated version of the $H$ function that contains only three terms:
\begin{equation}\label{eqn:approxh} H(x)=b_{1}\sin(x)+b_{2}\sin(2x)+a_{1}\cos(x)\end{equation}
Unless otherwise stated, we typically take $b_1=1$ and $b_2=-.75$, giving zeros of $H_{odd}$ at $\phi=0,\pi,\pm \cos^{-1}(2/3)\approx \pm 0.847.$  We remark that this particular form of $H_{odd}(\phi)$ arises in the bead on a hoop instability \cite{strogatz}(Chapt 3.5) and in a model for the coordination of finger tapping (\cite{kelso}). If $b_2<0$ is fixed and negative, then as $b_1$ decreases from a large positive value, the synchronous solution loses stability (at $b_1=-2b_2$) and a branch of intermediate stable phase-locked solutions bifurcates. This branch remains stable until $b_1$ becomes sufficiently negative $b_1<2b_2$ whereupon the anti-phase solution is stable.

\section{Waves in Large Networks}\label{Chapter4}

This section is a study of wave behavior in both chains and two dimensional arrays of neurons with nearest neighbor coupling in regimes where there is an intermediate stable phase-locked state for pair-wise coupling. 
Primarily, we study phase models which use the interaction functions (or approximations of them) derived from the Wang-Buszaki model.
Our models may be relevant to patterns of wave activity in the neonatal rat. Peinado et al. was able to observe wave activity in gap-junction coupled interneurons in rat neocortex prior to day twelve of development\cite{peinado}. Furthermore, he was able to enhance these waves by applying halothane and picrotoxin. Picrotoxin blocks inhibitory synapses while halothane reduces the potassium conductance. In general, he observed that the reduction in potassium directly led to the formation of waves. Since our intermediate stable phase-locked states can occur in gap junction coupled neurons by reducing the potassium conductance, this may be experimental evidence that this effect plays a role in wave formation in a two-dimensional network.\par
\indent We focus on a specific type of solution to the phase equations known as anti-waves. Anti-waves were first studied in two papers by Ermentrout and Kopell\cite{chains}\cite{ipp}. Similar phenomena have been examined by Strogatz et al., (uniformly twisted waves) and Blasius et al., (1D quasiregular concentric waves)\cite{sync}\cite{blasius}    
\footnote{These are not to be confused with fractured waves studied by Kopell (also known as s-waves) \cite{somers}. }.
Anti-waves consist of waves either initiated at the ends and colliding in the middle or waves initiating from the middle and terminating at the ends.  The latter are, in a sense, equivalent to one-dimensional target patterns. The wavenumber for these waves (the spatial gradient of the phase) shows an abrupt change of sign which we will call a kink.  
 Similar phase waves have been demonstrated in mechanical systems\cite{PhysRevLett.68.1730}. Anti-waves have been experimentally observed in the spinal cords of dogfish \cite{dogfish} and may well be present in other biological tissue. For instance, similar patterns of electrical activity have been observed in the muscle of the colon of a cat \cite{cat}. Central pattern generators in the fins of electric fish have also been known to produce anti-waves\cite{ipp}. These animals are able to produce a variety of complex waves in which the ``kink'' or lead oscillator in the wave is able to shift\cite{ipp}. 

We want to stress that our mechanism for generating anti-waves is fundamentally different than previous papers. Furthermore, our anti-waves can form anywhere along the chain and are much more robust to perturbations than in previous models\cite{chains}. 
The rest of this paper is organized into several sections. We begin in section \ref{models} by discussing the basic phase models and boundary conditions. This is followed by an analysis of both an ordinary traveling wave (\ref{waves1}) and anti-wave solutions (\ref{waves2}) . In section \ref{prob}, we demonstrate that the probability of obtaining a particular solution depends on the relative contribution of the even component of the interaction function. We show that starting from the anti-wave solution, if the even component is sufficiently large, perturbations initiated at one end of the chain can propagate down the chain and shift the position of the kink. Finally, in \ref{2D} we demonstrate that this analysis can be extended to higher dimensions and that a variety of anti-wave patterns are possible in a two dimensional oscillator arrays.

\subsection{Models And Boundary Conditions}\label{models}
The models we consider were introduced by Kopell and Ermentrout in a 1986 paper\cite{chains}.
These models primarily describe networks of neurons with nearest neighbor coupling. 
In analyzing these equations we apply two types of boundary conditions: periodic boundary conditions and non-reflecting boundary conditions. For a system of $N+1$ neurons with periodic boundary conditions, the system of phase equations may be written:  
\begin{eqnarray}\label{eqn:phase}
\dot{\theta_{1}}&=&\omega_{1}+H_{L}(\theta_{N+1}-\theta_{1})+H_{R}(\theta_{2}-\theta_{1})\nonumber\\
\dot{\theta_{2}}&=&\omega_{2}+H_{L}(\theta_{1}-\theta_{2})+H_{R}(\theta_{3}-\theta_{2})\nonumber\\
&\vdots&\nonumber\\
\dot{\theta}_{N+1}&=&\omega_{N+1}+H_{L}(\theta_{N}-\theta_{N+1})+H_{R}(\theta_{1}-\theta_{N+1}).\nonumber\\
\end{eqnarray}
In these equations, the $\omega_{i}$ represents the natural frequencies of the oscillators.
We denote the coupling in the two possible directions as $H_{L}(\phi)$ and $H_{R}(\phi)$.  In general, we assume that the coupling is isotropic, so that: $H_{R}(\phi)=H_{L}(\phi)=H(\phi)$.
Ultimately, we are interested in phase-locking behavior, thus we make the change of variables: $\phi_{j}=\phi_{j+1}-\phi_{j}$.
 This results in a system of $N$ phase equations:

\begin{eqnarray}\label{eqn:phasediff}
\dot{\phi_{1}}&=&\Delta \omega_{1}+H(\phi_{2})+H(-\phi_{1})-H(\sum^{N}_{j=1}\phi_{j})-H(\phi_{1})\nonumber\\
\dot{\phi_{2}}&=&\Delta \omega_{2}+H(\phi_{3})+H(-\phi_{2})-H(-\phi_{1})-H(\phi_{2})\nonumber\\
\dot{\phi_{j}}&=&\Delta \omega_{j}+H(\phi_{j+1})+H(-\phi_{j})-H(-\phi_{j-1})-H(\phi_{j})\nonumber\\
\vdots\nonumber\\
\vdots\nonumber\\
\dot{\phi_{N}}&=&\Delta \omega_{1}+H(-\phi_{N})+H(-\sum^{N}_{j=1}\phi_{j})-H(-\phi_{N-1})-H(\phi_{N}).\nonumber\\
\end{eqnarray}
In these equations $\Delta \omega_{i}$ is the frequency gradient between oscillators. In most of our simulations, we assume identical frequencies so $\Delta \omega_{i}=0$\\
The second type of boundary condition that we use is a variation of what  are known as non-reflecting boundary conditions \cite{KevinW}. Non-reflecting boundary conditions are implemented in order to attempt to eliminate reflections and to ``trick" the neurons at the ends of the chains, neuron $1$ and neuron $N+1$, into behaving as though the chain is infinite. If one were to think of the chain as being a continuous system, then the boundary conditions are simply a statement that $\frac{d\phi(x,t)}{dx}=0$ when evaluated at the ends of the chain.
There is a precedent for using such boundary conditions in nonlinear oscillator problems, for an example, see \cite{PhysRevLett.68.1730} .
Applying these boundary conditions to our phase equations, we have\footnote{To see that this is consistent with the preceding statement, start with the $H(-\phi_{j-1})=H(\phi_{j})$, using a Taylor series to expand out each side and evaluating at the wave solution, we have the expression:
\begin{eqnarray}
H'(-\phi)\frac{d\phi}{dx}=0
\end{eqnarray} 
 Since $H'(\phi)\neq 0$, then it must be that $(\frac{d\phi}{dx}|_{x=0})=0$.}:
\begin{eqnarray}
\theta_{0}&=&\theta_{2}\nonumber\\
\theta_{N+1}&=&\theta_{N-1},
\end{eqnarray}
 and our phase equations become:
\begin{eqnarray}\label{eqn:reflectedbc}
\dot{\phi_{1}}&=&\Delta \omega_{1}+H(\phi_{2})+H(-\phi_{1})-2H(\phi_{1})\nonumber\\
\dot{\phi_{2}}&=&\Delta \omega_{2}+H(\phi_{3})+H(-\phi_{2})-H(-\phi_{1})-H(\phi_{2})\nonumber\\
\dot{\phi_{j}}&=&\Delta \omega_{j}+H(\phi_{j+1})+H(-\phi_{j})-H(-\phi_{j-1})-H(\phi_{j})\nonumber\\
\vdots\nonumber\\
\vdots\nonumber\\
\dot{\phi_{N}}&=&\Delta \omega_{N}+2H(-\phi_{N})-H(-\phi_{N-1})-H(\phi_{N}).\nonumber\\
\end{eqnarray}
 
\subsection{Traveling Waves In Chains Of Coupled Oscillators}\label{waves1}
\begin{figure}
\centering
\includegraphics[width=6in, height=4in]{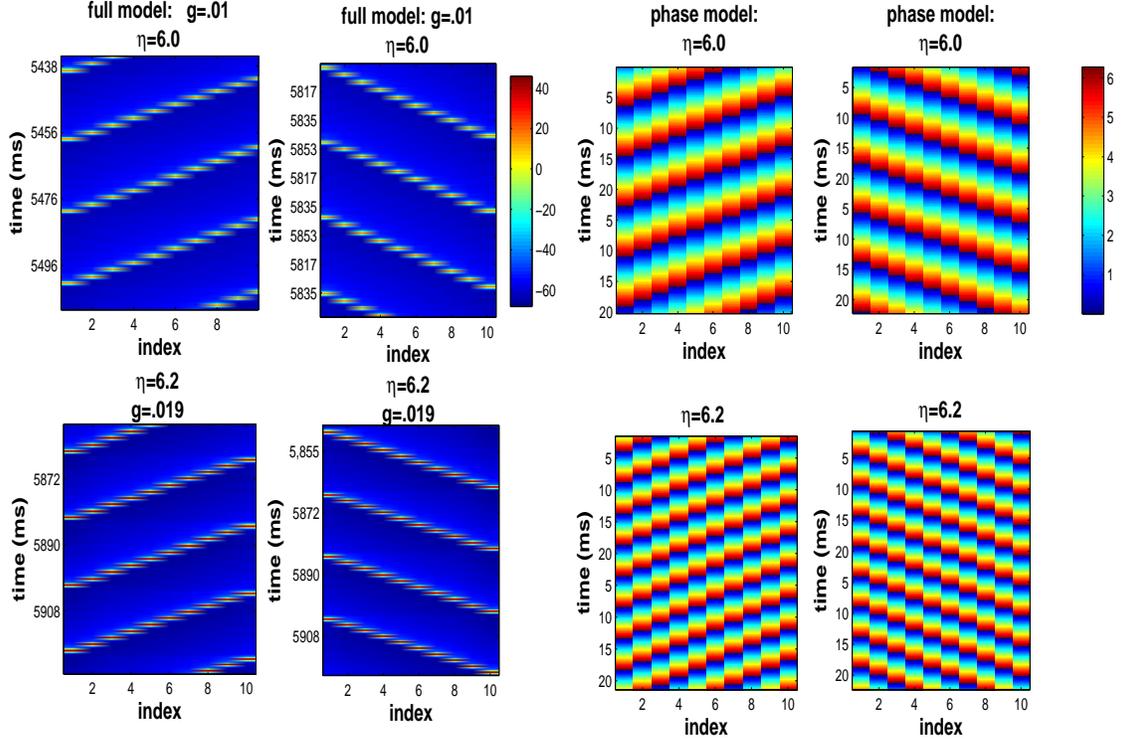}
\caption{Examples of traveling waves in both the Wang-Buszaki model (left) and the phase model (right) corresponding to two different values of the temperature dependent time constant $\eta$. The phase model reproduces the dynamics of the full model. Furthermore, it is clear from the phase model, that as one increases the constant $\eta$, the wavelength of the traveling waves decreases. Coupling strength in the full model is small: It must be on the order of $g=0.01\hspace{.1in}\mu S/cm^2$ to ensure phase-locking. }
\label{fig:tw}
\end{figure}    
The wave solution for the traveling wave can be written:
\begin{eqnarray}\label{eq:tw}
\phi_{j}= j\phi^* +\Delta\omega t,
\end{eqnarray}
where $j$ is the index of the oscillator and $\phi^*$ is the phase shift between adjacent oscillators.
Substituting this solution into the equations with non-reflecting boundary conditions, we see that \ref{eq:tw}
is a solution provided that $H(\phi^*)=H(-\phi^*)$. Since our interaction function has both odd and even components, this statement is true only if $\phi^*$ is the root of $H(\phi)_{odd}$. Thus, the stable phase-locked states for pair-wise symmetrically coupled oscillators determine the wavenumber, $\phi^*$. In a continuum limit, we can consider $\theta(x,t)$ to be a function of both position and time. If we take the derivative of  $\theta(x,t)$ with respect to $t$ we obtain the expression\cite{chains}:
\begin{eqnarray}\label{eqn:continuum}
\frac{d\theta}{dt}+\frac{\partial\theta}{\partial x}\frac{\partial x}{\partial t}=0,
\end{eqnarray}
which is equivalent to:
\begin{eqnarray}
\omega+\phi^* v_{\theta}=0\nonumber\\
v_{\theta}=\frac{\omega}{\phi^*}.
\end{eqnarray}
This is an expression for the phase velocity of the wave. Therefore, we see that we may identify the wavenumber of the system as: 
$k=\phi^*$. The stable phase-locked state between pairs of oscillators defines the wavenumber of a traveling wave. In the last section it was demonstrated that as we vary constants $\eta$ and $g_{k}$ in the Wang-Buszaki model, the stable fixed point changes. This translates to a change in wavelength in a chain of neurons.  Figure \ref{fig:tw} shows a comparison between the phase model and full model for a variety of $g_{k}$ and $\eta$. The four panels on the left correspond to the full model. The panels on the right correspond to the phase model. The phase model quantitatively reproduces the dynamics of the full model. The phase model clearly demonstrates that the wavenumber increases for increasing $\eta$. Coupling strength in the full model is small: it must be on the order of $0.01$ to get agreement with the phase models. Stronger coupling results in only synchronous dynamics.

\subsection{Anti-Waves In Chains of Coupled Oscillators}\label{waves2}
Anti-waves have been studied by Ermentrout and Kopell in two separate publications \cite{chains}\cite{ipp}. 
The previous mechanisms for generating anti-waves rely on extremely long chains (essentially infinite) or chains with distal connections. \par
\begin{figure}
\centering
\includegraphics[width=5.5in]{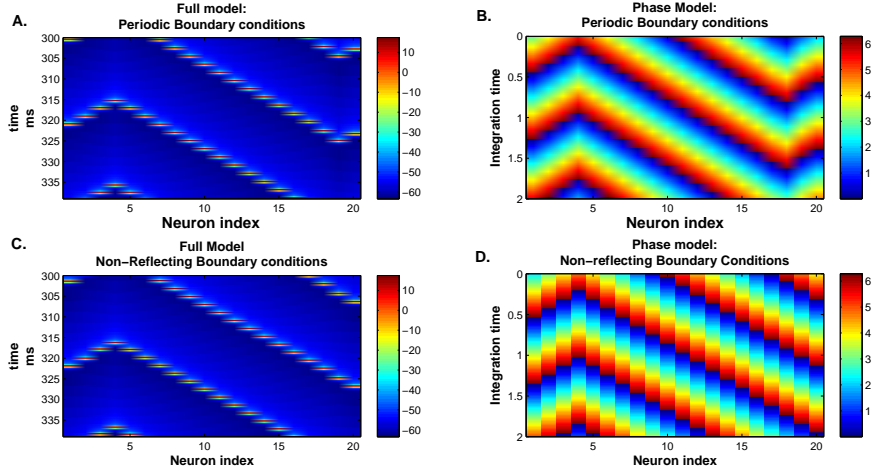}
\caption{Four examples of anti-waves in both rings and chains of oscillators computed with $\eta=6.0$.
{\bf{A.}} Is a wave in a chain of Wang-Buszaki neurons with periodic boundary conditions. {\bf{B.}} Is a wave in a chain of phase oscillators with periodic boundary conditions. { \bf{C.}} Is a wave in a ring of Wang- Buszaki neurons with non-reflecting boundary conditions (full model). {\bf{D.}} Is the phase model reduction of panel C: it is a wave in a chain of phase oscillators with non-reflecting boundary conditions.}
\label{fig:fwaves}
\end{figure}

\indent Assuming an isotropic chain with no gradient in the natural frequencies, the intermediate stable phase-locked state defined by $H(\phi^*)_{odd}=0$ will generate traveling waves. If the fixed point $\phi^*$ is identified as the wavenumber $k$, then the one kink anti-wave solution can be written:
\begin{eqnarray}\label{eqn:shock}
\phi_{j}= k j+\omega t \hspace*{1in}  j < j^*\nonumber\\
\phi_{j}=- k j+\omega t \hspace*{1in}  j\geq j^*.
\end{eqnarray}
In these equations $j^*$ represents the position of the kink.
By substituting the above expression into \ref{eqn:reflectedbc} we see again that the anti-wave is a solution provided that $H(k)=H(-k)$ ($k$ is the root of $H_{odd}(k)$) .
Figure \ref{fig:fwaves} demonstrates examples of anti-waves obtained in the Wang-Buszaki model compared with a phase model. Figures \ref{fig:fwaves}A and \ref{fig:fwaves}C are waves generated in the full model with non-periodic and non-reflecting boundary conditions, respectively. Figures \ref{fig:fwaves}B and \ref{fig:fwaves}D are the equivalent phase models.
\subsection{Obtaining Different Wave solutions from Random Initial Conditions}\label{prob}
\begin{figure}
\centering
\includegraphics[width=5in, height=4in]{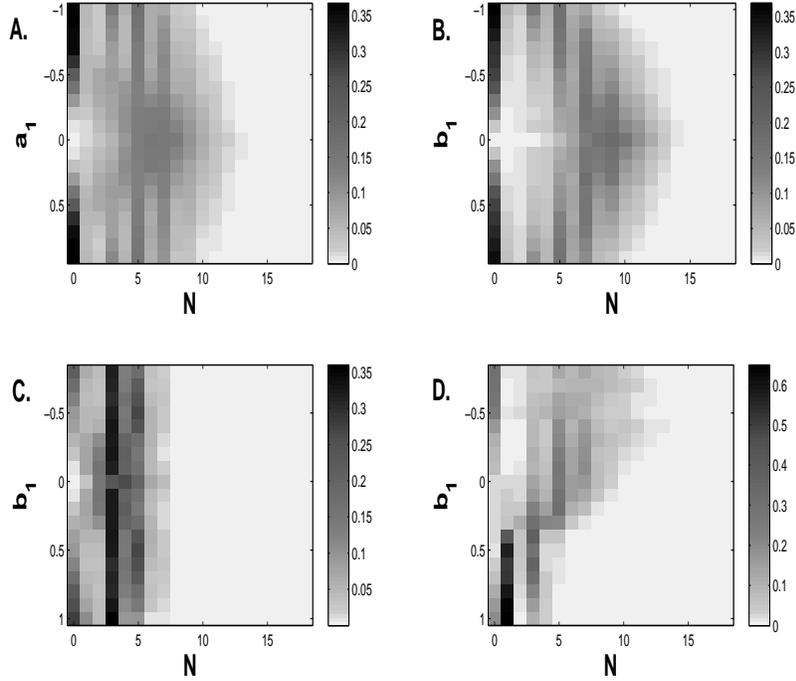}
\caption{Probability of obtaining different solutions of equation (\ref{eqn:reflectedbc}) for twenty oscillators as a function of the Fourier coefficients of the interaction function. $N$ designates the number of shocks in an anti-wave. $N=0$ corresponds to a traveling wave. {\bf{A.}} The probability distribution  with $H(\phi)=a_{1}\cos(\phi)+b_{1}\sin(\phi)+0.75\sin(2\phi)$ as a function of $a_{1}$ with $b_{1}=1$. {\bf{B.}} The probability distribution with $a_{1}=1$ and varying $b_{1}$. {\bf{C.}} The probability distribution calculated with an interaction function: $H(\phi)=-3\cos(\phi)-.92\cos(2\phi)+b_{1}\sin(\phi)-0.75\sin(2\phi)$. {\bf{D.}} The probability distribution calculated using the interaction function:  $H(\phi)=-3\cos(\phi)+b_{1}\sin(\phi)-0.75\sin(2\phi)$. For each parameter value the equations were solved from 10000 random initial conditions.}
\label{fig:prob1}
\end{figure}    
In order to see the variety of anti-waves, we will start chains of oscillators with random initial phases to estimate the basins of attraction \cite{sync}.
The interaction function we use (\ref{eqn:approxh}) allows any number of shocks or kinks in the anti-wave, constrained only by the length of the chain. However, if we start the equations from random initial conditions, the probability of getting a certain number of shocks varies as we change the size of the Fourier components. The main point is that even if the odd portion of the interaction function allows for a multiple shock solution, the probability of the system converging to this solution from random initial conditions may be extremely low and is determined by the magnitude of both the even and first odd Fourier modes. 
Figure \ref{fig:prob1} shows the  probability distributions of obtaining various anti-wave and traveling wave solutions as a function of the magnitudes of both the even and odd Fourier terms of the interaction function. Panel \ref{fig:prob1}A is a plot of the probability of obtaining either an $N$-shock anti-wave solution or a traveling wave ($0-$shock) solution as a function of $a_{1}$. From the plot, we see that the probability of obtaining a traveling wave solution approaches zero as $a_{1}\rightarrow 0$   and the probability distribution shifts towards $N=6$. As $a_{1}$ increases towards $1$, the probability of obtaining the traveling wave solution increases until it is the most probable state. Panel \ref{fig:prob1}B is a plot of the probability distribution as a function of $b_{1}$ with $a_{1}=1$.  Panel \ref{fig:prob1}B shows a trend similar to \ref{fig:prob1}A. For $b_{1}=0$ the probability of obtaining a traveling wave solution from random initial conditions is close to $0$. The solution with the maximal probability corresponds to an anti-wave with $N=9$ shocks or kinks. As $b_{1}$ is increased towards $1$ or decreased towards $-1$ the probability distribution shifts towards $N=0$. That is, solutions with fewer kinks become more probable.
Panels \ref{fig:prob1}C and \ref{fig:prob1}D demonstrate the effect of higher order even terms. Panel \ref{fig:prob1}D demonstrates that using the interaction function with two even terms ($H(\phi)=\cos(\phi)+b_{1}\sin(\phi)-\frac{3}{4}\sin(2\phi)$) results in a probability distribution in which the traveling wave solution is the most probable solution for $b_{1}\rightarrow1$ and the 1 kink anti-wave solution is the most probable for $b_{1}\rightarrow -1$. If we do not include this second order term, as demonstrated in panel \ref{fig:prob1}C the most probable solution corresponds to an ($N=4$ shocks) anti-wave.  The main point of this is that, even though only the odd Fourier modes determine the solutions to equations \ref{eqn:phasediff} and \ref{eqn:reflectedbc}, the even Fourier modes can drastically affect the basin of attraction of those solutions.

\subsection{Moving The Shock Position With Impulses}
\begin{figure}
\centering
\includegraphics[width=5in, height=2.5in]{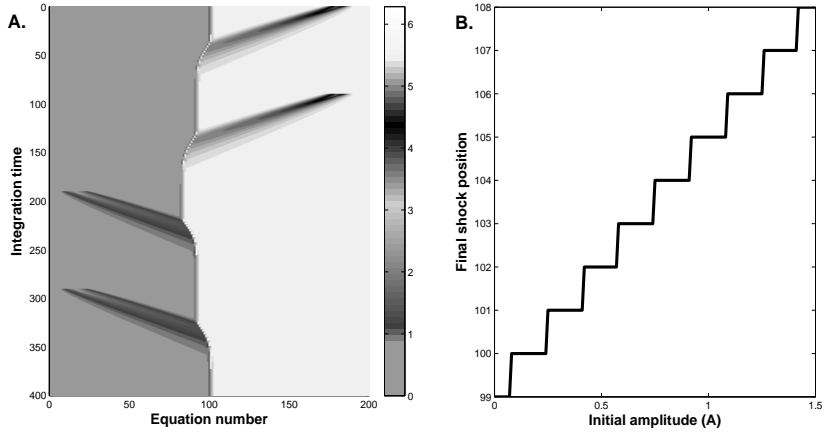}
\caption{{\bf{A.}} Compactons initiated from either side of the chain can shift the phase shock in the anti-wave back and forth. The interaction function used was: $H(\phi)=\cos(\phi)+\sin(\phi)-0.75\sin(2\phi)$ 
{\bf{B.}} The shift in the kink for varying initial compacton amplitudes.}  
\label{fig:compacton}\end{figure}    

Electric fish have been observed to produce complex anti-wave type patterns\cite{ipp}. What is more, the lead oscillator  or (kink) in these anti-waves has been observed to be able to shift position. We present a mechanism which shifts the position of the kink/shock in the anti-wave without changing its shape.
Over the last fifteen years, Pikovsky and Rosenau have written several papers studying waves in oscillator lattices with purely even coupling. They showed that such interaction functions can be derived from networks of Josephson junctions and Van-der Pol oscillators\cite{compacton1}\cite{compacton2}\cite{compacton3}. Waves in these networks take the form of solitary pulses which retain their shape. Pikovsky and Rosenau have coined these waves ``compactons''\cite{compacton1}\cite{compacton2}\cite{compacton3}. Collisions between different compactons have been studied in detail. Nothing, however, has written about compactons colliding with a phase boundary (like the phase shock in the anti-wave). 
 Compacton-like pulses are possible not only in systems with a purely even interaction function, but they can be observed in systems with odd terms, provided the even component is large enough\cite{compacton3}. The odd portion of the interaction function acts to dissipate the initial pulse, but a compacton-like wave may still travel large distances even with a substantial odd term. Compactons are possible in systems in which the interaction function generates anti-waves. For instance, we consider a system of 200 phase-difference equations with non-reflecting boundary conditions. The interaction function used is $H(\phi)=\cos(\phi)+\sin(\phi)-.75\sin(2\phi)$. 
The initial conditions are the one kink anti-wave solution with an extra pulse stimulus applied at the end.
Thus we have the one kink/shock solution:
\begin{eqnarray}\label{eqn:compacton1}
\phi_{j}(0)&=&k \hspace{2.1in} j<\frac{N}{2}   \nonumber\\
\phi_{j}(0)&=&-k \hspace{2.in} j\geq \frac{N}{2},   \nonumber\\
\end{eqnarray}
with the addition of a pulse:
\begin{eqnarray}\label{eqn:compacton2}
\phi_{j}(0)=k+\frac{A}{2}(1+\cos((j-x_0)\pi/\sigma))  \hspace{.25in} |j-x_0|<\sigma.
\end{eqnarray}
Here $A$ is the amplitude, $x_0$ is the position of the pulse and $\sigma$ is the width of the pulse. In Figure \ref{fig:compacton} the pulse width used is $\sigma=10$ and $|k|=.84106$.
This pulse is the form used by Pikovsky et al. to generate compactons, but other initial conditions may work as well\cite{compacton3}.
Figure \ref{fig:compacton}C. shows multiple compactons traveling on top of an anti-wave and colliding with the shock located at $N=100$. The anti-wave is composed of the stationary white and grey regions.
The grey region of the plot corresponds to the solution $\phi=.841$ whereas the white region corresponds to: $\phi=5.44$. Upon collision, the shock shifts but retains its shape. In this manner, multiple pulses initiated at the ends of the chain may be used to shift the shock back and forth. 
Figure \ref{fig:compacton}B. is a plot of the shift in the kink as a function of the initial compacton amplitude. The maximum shift obtained is approximately 9 sites. If $A>1.5$ radians, larger amplitude pulses do not necessarily provide larger position shifts. If the pulse is too large, it will destroy the ``perfect kink'' solution. Thus, near the supercritical pitchfork bifurcation, not only can the shock of an anti-wave form anywhere along the the chain but precisely because of this property, it can be shifted around by an impulse (compacton). In this way, the additional even and odd Fourier terms produce a central pattern generator which is malleable. Perhaps this is a mechanism by which the hindbrain of a fish could send impulses to the rest of its spinal cord to modify the fish's swimming pattern.
\section{Stability Analysis}
In analyzing the stability of anti-wave solutions, one of the main questions we want to address is how the relative contributions of each Fourier mode contributes to the stability of the solution. Additionally, we want to examine the importance of other parameters in the model, such as the length of the chain and the position of the kink. In these chain models there  are four basic wave solutions which are of interest. Two of the solutions correspond to a traveling wave in either direction (left to right or right to left) and two correspond to anti-waves. The first anti-wave solution describes a wave emanating from the center of the chain and propagating in both directions outward. The second anti-wave solution corresponds to two waves emanating from the edges and colliding in the center of the chain. 
Using our simplified interaction function we begin by analyzing the simplest (shortest) chain possible, the three neuron system.
 The three phase equations (\ref{eqn:3chain}) describing the neurons can be condensed to two by a change of variables. Linearizing these equations (\ref{nullclines}) about the anti-wave solution results in a $2$x$2$ Jacobian. Thus, the problem is simple enough so that we can solve for the eigenvalues as a function of $a_{1}$  and subsequently show where and how the anti-wave solution loses stability. Once we have proved stability for this simple case, we discuss longer chains and specifically, we analyze the effects of the magnitude of the even component on the stability of various anti-wave solutions.

\subsection{Stability Analysis Of The Three Oscillator System}
\begin{figure}
\centering
\includegraphics[width=5in]{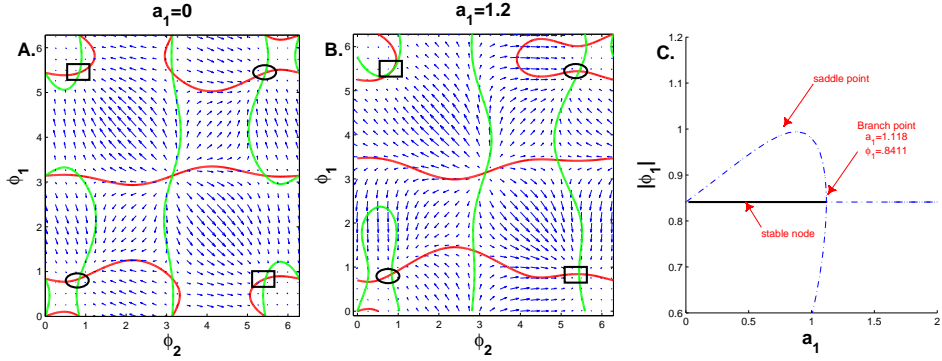}
\caption{Phase portrait of the two-dimensional system, (\ref{nullclines}) where $H(\phi)=a_1\cos(\phi)+\sin\phi-0.75\sin 2\phi$. The $\phi_1$ ($\phi_2$) nullcline is in red(green). 
 {\bf{A.}} The nullclines and the flow for $a_{1}=0$. The fixed points corresponding to anti-waves are enclosed with boxes. The fixed points corresponding to traveling waves are circled. {\bf{B.}} The nullclines and the flow for $a_{1}=1.2$.  {\bf{C.}} Bifurcation diagram computed with AUTO: The anti-wave solution loses stability near $a_{1}=1.118$.  In this plot, the stable solution is represented by a solid line whereas the unstable solution is represented by the dotted lines. 
 }
\label{fig:null}
\end{figure}

Our equations for the system with non-reflecting boundary conditions are:
\begin{eqnarray}\label{eqn:3chain}
\dot{\theta_{1}}=2H(\theta_{2}-\theta_{1})\nonumber\\
\dot{\theta_{2}}=H(\theta_{3}-\theta_{2})+H(\theta_{1}-\theta_{2})\nonumber\\
\dot{\theta_{3}}=2H(\theta_{2}-\theta_{3}).
\end{eqnarray}
We then write them in terms of their phase differences:
\begin{eqnarray}\label{nullclines}
\dot{\phi_{1}}=2H(-\phi_{2})-H(-\phi_{1})-H(\phi_{2})\nonumber\\
\dot{\phi_{2}}=H(-\phi_{1})+H(\phi_{2})-2H(\phi_{1}).
\end{eqnarray}
Note that these equations are invariant under a reflection:
\begin{math}
\phi_{1}\rightarrow -\phi_{2}\\
\phi_{2}\rightarrow -\phi_{1}\\ 
\end{math}
Figure \ref{fig:null} is a plot of the nullclines of the system \ref{nullclines} for various values of $a_{1}$ using $b_1=1$ and $b_2=-.75$.
There are two anti-wave solutions indicated by the boxes and two traveling wave solutions which are circled. One can see that when there is no even component the solutions to the system possesses perfect reflection symmetry. As soon as $a_{1}$ is nonzero, the system loses this symmetry.  
We want to analyze the stability of the anti-wave analytically. Choosing $\phi_{1}=k$ and $\phi_{2}=-k$, we linearize our equations about this system and write down the Jacobian as follows:
\begin{eqnarray}\label{eqn:3jacobian}
M_{0}=\left(\begin{array}{cc}
H'(-k)&-2H'(k)-H'(-k)\\
-H'(-k)-2H'(k)&H'(-k)\\
 \end{array} \right),
\end{eqnarray}
Solving for the eigenvalues of this expression, we have
\begin{eqnarray}\label{eqn:eigenvalues}
\lambda_{1,2}=-2H'(k),-2H'(-k)-2H'(k).
\end{eqnarray}
 As long as the derivative of these H functions evaluated at this solution is positive, then the solution will be stable.
Substituting the simplified interaction function (\ref{simple}) and its derivative into equation \ref{eqn:eigenvalues}.
results in the eigenvalue expressions: equation \ref{eqn:seigenvalues}.
\begin{eqnarray}\label{simple}
H(\phi)&=&b_{1}\sin(\phi)+b_{2}\sin(2\phi)+a_{1}\cos(\phi)\nonumber\\
H'(\phi)&=&b_{1}\cos(\phi)+2b_{2}\cos(2\phi)-a_{1}\sin(\phi)\nonumber\\
\end{eqnarray}

\begin{eqnarray}\label{eqn:seigenvalues}
\lambda_{1}=-2(b_{1}\cos(k)+2b_{2}\cos(2k)-a_{1}\sin(x))\nonumber\\
\lambda_{2}=-4(b_{1}\cos(k)+2b_{2}\cos(2k))
\end{eqnarray}

At the critical value of the parameter  $a_{critical}=1.1$,  the first eigenvalue vanishes. For larger values of $a_1$, the fixed point becomes unstable. The mirror symmetry of the equations, along with the bifurcation diagram in Figure \ref{fig:null}, suggest that this is a subcritcal pitchfork bifurcation. This is verified in \cite{Alex} where we explicity calculate the normal form equations.

\subsection{Stability Analysis For The Traveling Wave}
Proving the stability of the traveling wave (with no kinks) is relatively straightforward.
We begin by writing down the equations for the ``jth" oscillator.
Since the system is nearest neighbor coupling, the general expression is:
\begin{eqnarray}\label{eqn:tw1}
\dot{\phi_{j}}=H(\phi_{j+1})+H(-\phi_{j})-H(-\phi_{j-1})-H(\phi_{j}).
\end{eqnarray}
Defining $a=H'(\phi^*)$, $b=H'(-\phi^*)$ and linearizing the equations about the wave solution, the equations for the phase are simply a discretized version of Laplace's equation:
\begin{eqnarray}\label{eqn:tw2}
a \phi_{j+1}-(a+b)\phi_{j}+b\phi_{j-1}=\lambda \phi_{j}.
\end{eqnarray}
We may solve for the equations by assuming a general solution:
\begin{eqnarray}\label{eqn:soln}
\phi_{j}=Ax^{j}_{1}+Bx^{j}_{2},
\end{eqnarray}
and invoking the boundary conditions:
$\phi_{0}=\phi_{1}$ and $\phi_{N-1}=\phi_{N}$
.Plugging our solution \ref{eqn:soln} into \ref{eqn:tw2}, we may solve for the eigenvalues of the system: 
\begin{eqnarray}\label{eq:eigenvalues}
\Re{\lambda}=2\sqrt{ba}\cos{\frac{\pi m}{N-1}}-(a+b).
\end{eqnarray}
Therefore the wave solution will always be stable provided that
\begin{eqnarray}
2\sqrt{ba}\leq a+b
\end{eqnarray}
Or alternately that $a,b>0$. Stability is lost for $a,b<0$.
\subsection{Stability for The Anti-Wave under More General Conditions}
Stability analysis of the anti-wave solutions may be proven using a combination of the Gershgorin Circle Theorem and numerically computing the eigenvalues of the linearized equations.
In these examples, we use non-reflecting boundary conditions. The argument will be identical for periodic boundary conditions. One starts by assuming a 1 shock solution and linearizing the phase the equations about this solution.
The Jacobian will have matrix elements of the form:
\begin{eqnarray}
a_{i,i}=-(H'(\phi_j)+H'(\phi_j)) \hspace{.25in} a_{i,i+1}=H'(\phi_{j+1}) \hspace{.25in} a_{i,i-1}=H'(\phi_{j-1}) 
\end{eqnarray} 
All other matrix elements are zero. Once more, define: $a=H'(\phi^*)$, $b=H'(-\phi^*)$.  If $l$ denotes the location of the kink, then the matrix elements corresponding to the kink are:
\begin{eqnarray}\label{eqn:gg1}
a_{l,l}=-(a+b) \hspace{.25in} a_{l,l+1}=b \hspace{.25in} a_{l,l-1}=b 
\end{eqnarray} 
or,
\begin{eqnarray}\label{eqn:gg2}
a_{l,l}=-(a+b) \hspace{.25in} a_{l,l+1}=a \hspace{.25in} a_{l,l-1}=a
\end{eqnarray} 
depending on the orientation of the shock. The Gershgorin Circle Theorem \footnote{Gershgorin Circle Theorem:
Let:$ A=[a_{ij}]$ be an arbitrary n x n matrix with elements that may be complex and let:\\
\begin{eqnarray}
\Lambda_{i}=\sum^{n}_{j=1,i\neq j}|a_{ij}|\hspace{.25in} for\hspace{.1in} i=1,2,...n
\end{eqnarray}
Then all of the eigenvalues $\lambda_{i}$ of$A$ lie in the union of n disks $\Gamma_{i}$ where:
\begin{eqnarray}
\Gamma_{i}:|\lambda-a_{ii}|\leq|\lambda_{i}\hspace{.25in} for\hspace{.1in} i=1,2,...n
\end{eqnarray} 
This wording of the Gershgorin Circle theorem was taken from:\\
Tables of Integrals, Series, and Products by I.S.Gradshteyn and I.M. Ryzhik\cite{integrals}. 
} tells us that all of the eigenvalues of the matrix lay in the union of disks centered at the diagonal elements of the matrix with radii less than the absolute value of the sum of the row entries not including the diagonal terms. If the Jacobian of the anti-wave is described by equation \ref{eqn:gg1} all eigenvalues will lie in the union of three disks: one centered at $-(a+b)$ with radius $a+b$, one centered at $-2a+b$ (corresponding to the ends of the chain) with radius $a$ and one centered at $-(a+b)$ with radius $2b$. If we assume that the even term $a_{1}$ is positive and increasing, then $H'(-\phi) \geq H'(\phi)$ 
\begin{eqnarray}
|b|\geq|a|,\nonumber\\
|2b|>|a+b|.
\end{eqnarray}
Thus the disks will extend beyond the origin, and we will not be able to say anything about stability using this theorem. On the other hand, if the solution is a shock oriented in the opposite direction, such as in equation \ref{eqn:gg2}, the disk corresponding to the equation at the shock is centered at $-(a+b)$ and extends out with radius $|2a|$.
The disk will always lie in the left half of the imaginary plane for $a_{1}$ positive. Therefore, this solution will always be stable.
\begin{figure}
\centering
\includegraphics[width=6.0in,height=3.0in]{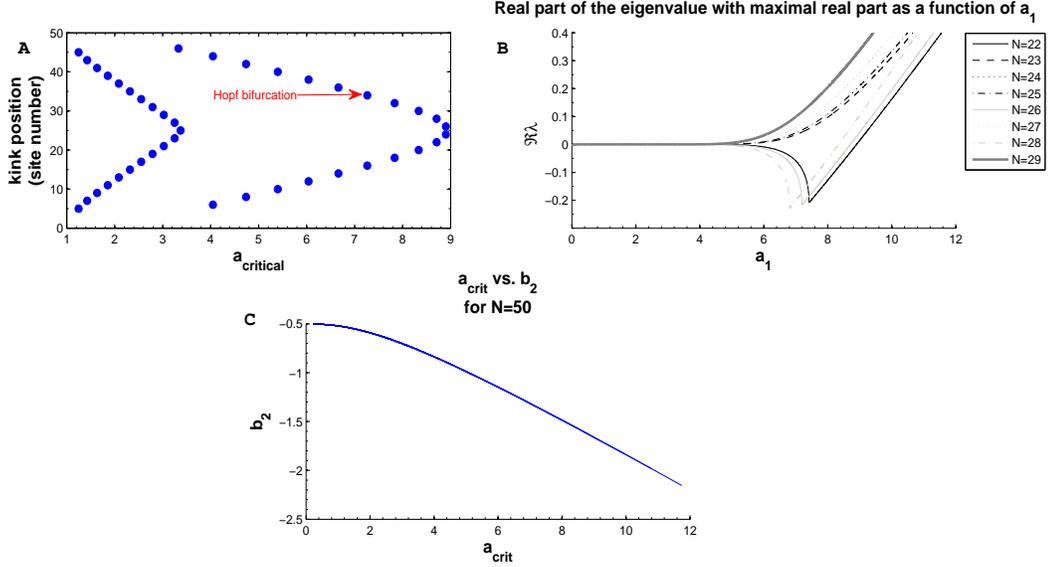}
\caption{{\bf{A.}} Critical value of $a_{1}$ as a function of shock position. {\bf{B.}} Eigenvalue with the maximal real part as a function of $a_{1}$ for different shock positions. $N$ denotes the location of the shock. {\bf{C.}} An interaction function with a more negative $b_{2}$ can also possesses a larger $a_{1}$ before the solution becomes unstable. }
\label{fig:stability1}
\end{figure}     
In the former case, in which the fractured wave is not necessarily stable, one must apply numerical methods to explicitly calculate the eigenvalues of the Jacobian. We are interested how variables such as the position of the shock and the length of the chain effect the stability of the solution as we vary the magnitude of the first even Fourier mode.
Again, we use the interaction function: $H(\phi)=b_1\sin(\phi)+b_2\sin(2\phi)+a_1\cos(\phi)$, where $b_1=1$ and $b_2=-.75$. 
We start by examining a fifty-one oscillator chain (described by fifty equations) in which we move the position of the shock. Figure \ref{fig:stability1}A shows the critical value of $a_{1}$ at which the shock solution loses stability as a function of shock position where the position varies between site $4$ and site $46$. The parameter $a_{critical}$ is determined by calculating the eigenvalues of the Jacobian for various values of $a_{1}$ and determining when the eigenvalue with the maximal real part becomes positive. Panel \ref{fig:stability1}B plots the real part of the eigenvalue with maximal real part as a function of $a_{1}$:
\begin{figure}
\centering
\includegraphics[width=4.0in]{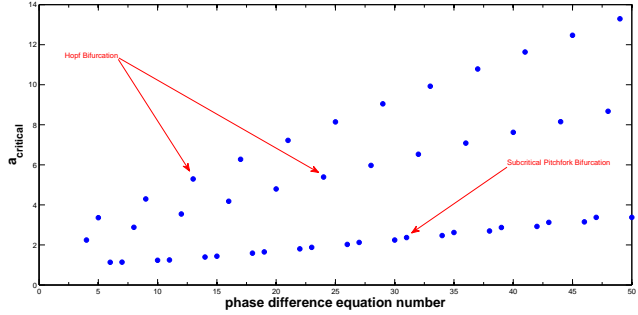}
\caption{Critical value of $a_{1}$ as a function of chain length where N represents the number of phase difference equations $\phi_{N}$. Depending on the length of the chain and position of the shock, the solution may lose stability in either a Hopf bifurcation or what is believed to be a subcritical pitchfork bifurcation. The shock is always located centrally: (e.g. for 101 oscillators, the shock is located at N=50. For an odd number of phase difference equations, the shock is position is obtained by dividing the number of equations by half and rounding up.)}
\label{fig:stability2}
\end{figure}    
as the position shifts, the eigenvalues lose stability at different values of $a_{1}$. The type of bifurcation by which they lose stability changes as well. 
For shocks located at even numbered sites, the system loses stability in a Hopf bifurcation as a complex conjugate pair of eigenvalues crosses the origin simultaneously. For shocks located at odd sites the system apparently loses stability in a sub-critical pitchfork bifurcation (This has not been rigorously proven). 
Figure \ref{fig:stability1}C is a plot of $a_{critical}$ for the eigenvalues of the fifty-one oscillator Jacobian linearized about solutions corresponding to varying values of $b_{2}$.  The plot clearly demonstrates that shock solutions corresponding to a larger value of $b_{2}$ can support a larger even component before becoming unstable. Figure \ref{fig:stability2} is a plot of $a_{critical}$ vs. the number of phase difference equations (number of oscillators in a chain). The shock is always located centrally: e.g. for 101 oscillators, the shock is located at $N=50$. For an odd number of phase difference equations, the shock is position is obtained by dividing the number of equations in half and rounding up.
 Figure \ref{fig:stability2} shows phenomena similar to Figure 10A. That is, a solution will lose stability for different $a_{1}$ dependent on the position of the shock in the solution as well as the number of oscillators in the chain. In the case that the shock is perfectly centered, we expect the solution to lose stability in a sub-critical pitchfork. Regardless of the position of the shock, even for relatively short chains the anti-wave solution will be stable for a relatively large even component, whose magnitude is at least as large as the first odd Fourier mode. The manner in which the solution loses stability and the size of the even component it can support depend on both the position of the kink/shock and the length of the chain. The Gershgorin circle theorem tells us that one of the anti-wave solutions will always be stable, no matter how long the chain.

\section{Pattern Formation in Two-Dimensional Arrays}\label{2D}

\begin{figure}
\centering
\includegraphics[width=5.0in]{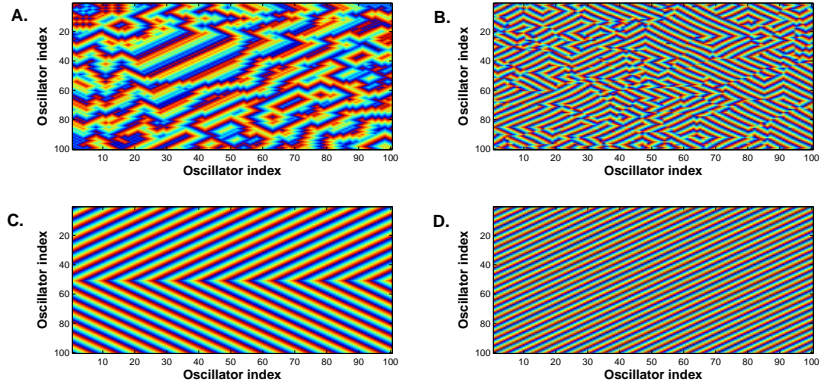}
\caption{ Examples of two-dimensional patterns. The horizontal and vertical axis are the oscillator indices.
{\bf{A.}} Snapshot of a slowly varying wave pattern obtained from compacton-like initial conditions: a 2-dimensional pulse is initiated in the upper left-hand corner of the array on a synchronous background. The interaction function used is $H(\phi)=\sin(\phi)-\cos(\phi)-.75\sin(2\phi)$ {\bf{ B.}} Stable stationary fractured pattern obtained from random initial conditions and interaction function:$-2\cos(\phi)-0.518\sin(\phi)-1.31\sin(2\phi)-.933\sin(3\phi)$ {\bf{ C.}} Stationary anti-wave generated with $H(\phi)=\cos(\phi)+\sin(\phi)-.75\sin(2\phi)$. {\bf{D.}} Stationary traveling wave generated with the same interaction function as {\bf{C.}} }
\label{examples}
\end{figure}

Anti-wave patterns can be observed in two-dimensional networks as well. Using nearest neighbor coupling,
the differential equation for a single oscillator is: 
\begin{eqnarray}\label{eqn:2darray}
\frac{d\theta_{x,y}}{dt}=H(\theta_{x+1,y}-\theta_{x,y})+H(\theta_{x,y+1}-\theta_{x,y})+H(\theta_{x,y-1}-\theta_{x,y})+H(\theta_{x-1,y}-\theta_{x,y}),
\end{eqnarray}
where $x,y$ are discrete indices describing the location of an oscillator.
These indices run from $1$ to $N$ where $N^2$ is the number of differential equations in the array. 
As in the one dimensional case, we use non-reflecting boundary conditions.
Examples of these patterns and the interaction functions used to generate them are are illustrated in Figure \ref{examples}. Patterns \ref{examples}B,\ref{examples}C and \ref{examples}D are stationary and numerically stable. Pattern \ref{examples}B is generated from random initial conditions. Patterns \ref{examples}C and \ref{examples}D are stable but have a very small basin of attraction. The initial conditions must be almost identical to the actual anti-wave or traveling wave solution.  

Analogous to the one-dimensional case, a plane wave may be represented by the solution:
\begin{eqnarray}
\theta_{xy}=k_{x}x+k_{y}y+\omega t.
\end{eqnarray} 
Whereas the shock solution may be written:
\begin{eqnarray}
\theta_{xy}=k_{x}x+k_{y}y \hspace*{.25in} x\leq x^*\nonumber\\
 \theta_{xy}=-k_{x}x+k_{y}y \hspace*{.25in} x> x^*. 
\end{eqnarray}
In these equations $x^*$ is the location of the shock.
A shock may be thought of as a discontinuous boundary between the two traveling waves. 
The fractured pattern in Figure \ref{examples}B is composed of many small waves of varying $k_{x}$ and $k_{y}$ which form shocks. 
Fractured patterns are generated with random initial conditions where the phases are chosen between $0$ and $2\pi$\footnote{All equations in this section were integrated with Euler's method. The step size used is $.01$.}  \footnote{The Fourier terms used may or may not be calculated from the full model. For instance, an interaction function with an artificially inflated second odd mode will produce an interesting fractured pattern, whereas many interaction functions computed from the full model will not.}.

\section{Conclusion}
Intermediate stable phase-locked states occur in a variety of neuron models and are capable of producing a pattern of wave-like activity known as an anti-wave. This type of activity was first observed in the dogfish spinal cord by Grillner \cite{dogfish} and it may occur in other networks as well\cite{cat}\cite{kim}.
The mechanism we use for generating these waves may have an experimental basis: Peinado et al. have shown that  modulating the potassium conductance of gap junction coupled neurons in the neonatal mouse neocortex leads to wave behavior \cite{peinado}.
 The mechanism for generating traveling waves and anti-waves is more flexible than past models, because it allows one to modulate the wavelength of the wave by adjusting properties such as the potassium conductance or temperature dependent time constant. It allows for a huge variety of phase-locked patterns in a chain or array of neurons. In the case of anti-wave generation, it does not require distal connections and the shock can form anywhere along the chain. The shock position can also be moved by colliding phase compactons. This may provide a mechanism by which an animal such as the dogfish could switch between different swimming patterns. More generally, near the bifurcation from in-phase synchrony, the interaction functions of the phase model have large higher order Fourier modes. This is in contrast to many oscillator models that were used in the past, which only include the lowest odd Fourier mode. The odd modes determine the value of the  stable-phase locked solution but the even modes affect the stability of the anti-wave solution. Varying the relative even component affects the basin of attraction of a particular solution, or the probability that the phases will converge to a particular anti-wave solution from random initial conditions. Finally, we note that this behavior is not only relevant to chains but to two dimensional arrays as well.  

\begin{appendix}
\section{The Wang-Buszaki Model}\label{model_eqns}
The parameters used in the Wang-Buszaki  model are: $v_{syn}=-60.5$mV,   $g_L=0.1 \mu$S, $v_L=-65$mV, $g_{Na}=35  \hspace{.1in} \mu S/cm^2$, $V_{Na}=55$mV, $g_{K}=9 \hspace{.1in}\mu S/cm^2$, $V_{K}=-90$mV, $a_{i0}=4$ $\tau_{i}=15$ms and $i_{0}=.63\hspace{.1in} nA/cm^2$ .
The nonlinearites mentioned in Section \ref{model} are given by the following expressions: 
\begin{eqnarray}
\alpha_m(V)=0.1(V+35.0)/(1.0-\exp(-(V+35.0)/10.0))\nonumber\\
\beta_m(V)=4.0\exp(-(V+60.0)/18.0)\nonumber\\
M_{\infty}(V)=\alpha_m(V)/(\alpha_m(V)+\beta_m(V))\nonumber\\  
\alpha_h(V)=0.07\exp(-(V+58.0)/20.0)\nonumber\\
\beta_h(V)=1.0/(1.0+\exp(-(V+28.0)/10.0))\nonumber\\  
H_{\infty}(V)=\alpha_h(V)/(\alpha_h(V)+\beta_h(V))\nonumber\\  
\alpha_n(V)=0.01(V+34.0)/(1.0-\exp(-(V+34.0)/10.00))\nonumber\\
\beta_n(V)=0.125\exp(-(V+44.0)/80.0)\nonumber\\
N_{\infty}(V)=\alpha_n(V)/(\alpha_n(V)+\beta_n(V))  
\end{eqnarray}
\end{appendix}

\bibliography{p3}

\begin{thebibliography}{10}%
\makeatletter
\providecommand \@ifxundefined [1]{%
 \ifx #1\undefined \expandafter \@firstoftwo
 \else \expandafter \@secondoftwo
\fi
}%
\providecommand \@ifnum [1]{%
 \ifnum #1\expandafter \@firstoftwo
 \else \expandafter \@secondoftwo
\fi
}%
\providecommand \enquote [1]{``#1''}%
\providecommand \bibnamefont  [1]{#1}%
\providecommand \bibfnamefont [1]{#1}%
\providecommand \citenamefont [1]{#1}%
\providecommand\href[0]{\@sanitize\@href}%
\providecommand\@href[1]{\endgroup\@@startlink{#1}\endgroup\@@href}%
\providecommand\@@href[1]{#1\@@endlink}%
\providecommand \@sanitize [0]{\begingroup\catcode`\&12\catcode`\#12\relax}%
\@ifxundefined \pdfoutput {\@firstoftwo}{%
 \@ifnum{\z@=\pdfoutput}{\@firstoftwo}{\@secondoftwo}%
}{%
 \providecommand\@@startlink[1]{\leavevmode\special{html:<a href="#1">}}%
 \providecommand\@@endlink[0]{\special{html:</a>}}%
}{%
 \providecommand\@@startlink[1]{%
  \leavevmode
  \pdfstartlink
   attr{/Border[0 0 1 ]/H/I/C[0 1 1]}%
   user{/Subtype/Link/A<</Type/Action/S/URI/URI(#1)>>}%
  \relax
 }%
 \providecommand\@@endlink[0]{\pdfendlink}%
}%
\providecommand \url  [0]{\begingroup\@sanitize \@url }%
\providecommand \@url [1]{\endgroup\@href {#1}{\urlprefix}}%
\providecommand \urlprefix [0]{URL }%
\providecommand \Eprint[0]{\href }%
\@ifxundefined \urlstyle {%
  \providecommand \doi [1]{doi:\discretionary{}{}{}#1}%
}{%
  \providecommand \doi [0]{doi:\discretionary{}{}{}\begingroup
  \urlstyle{rm}\Url }%
}%
\providecommand \doibase [0]{http://dx.doi.org/}%
\providecommand \Doi[1]{\href{\doibase#1}}%
\providecommand \bibAnnote [3]{%
  \BibitemShut{#1}%
  \begin{quotation}\noindent
    \textsc{Key:}\ #2\\\textsc{Annotation:}\ #3%
  \end{quotation}%
}%
\providecommand \bibAnnoteFile [2]{%
  \IfFileExists{#2}{\bibAnnote {#1} {#2} {\input{#2}}}{}%
}%
\providecommand \typeout [0]{\immediate \write \m@ne }%
\providecommand \selectlanguage [0]{\@gobble}%
\providecommand \bibinfo [0]{\@secondoftwo}%
\providecommand \bibfield [0]{\@secondoftwo}%
\providecommand \translation [1]{[#1]}%
\providecommand \BibitemOpen[0]{}%
\providecommand \bibitemStop [0]{}%
\providecommand \bibitemNoStop [0]{.\EOS\space}%
\providecommand \EOS [0]{\spacefactor3000\relax}%
\providecommand \BibitemShut [1]{\csname bibitem#1\endcsname}%
\bibitem{grillner}%
  \BibitemOpen
  \bibfield{author}{%
  \bibinfo {author} {\bibfnamefont{S.}~\bibnamefont{Grillner}},\ }%
  \bibfield{journal}{%
  \Doi{10.1126/science.3975635}{\bibinfo {journal} {Science}}\ }%
  \textbf{\bibinfo {volume} {228}},\ \bibinfo {pages} {143} (\bibinfo {year}
  {1985}),\
  \Eprint{http://arxiv.org/abs/http://www.sciencemag.org/content/228/4696/143.%
full.pdf}{http://www.sciencemag.org/content/228/4696/143.full.pdf},\
  \url{http://www.sciencemag.org/content/228/4696/143.abstract}%
  \bibAnnoteFile{NoStop}{grillner}%
\bibitem{izhikevich}%
  \BibitemOpen
  \bibfield{author}{%
  \bibinfo {author} {\bibfnamefont{E.}~\bibnamefont{Izhikevich}},\ }%
  \emph{\bibinfo {title} {Dynamical Systems in Neuroscience: The Geometry of
  Excitabilty and Bursting}}\ (\bibinfo {publisher} {MIT press: Cambridge,
  Massachusetts},\ \bibinfo {year} {2007})%
  \bibAnnoteFile{NoStop}{izhikevich}%
\bibitem{buszaki}%
  \BibitemOpen
  \bibfield{author}{%
  \bibinfo {author} {\bibfnamefont{G.}~\bibnamefont{Buzsáki}}\ and\ \bibinfo
  {author} {\bibfnamefont{A.}~\bibnamefont{Draguhn}},\ }%
  \bibfield{journal}{%
  \Doi{10.1126/science.1099745}{\bibinfo {journal} {Science}}\ }%
  \textbf{\bibinfo {volume} {304}},\ \bibinfo {pages} {1926} (\bibinfo {year}
  {2004}),\
  \Eprint{http://arxiv.org/abs/http://www.sciencemag.org/content/304/5679/1926%
.full.pdf}{http://www.sciencemag.org/content/304/5679/1926.full.pdf},\
  \url{http://www.sciencemag.org/content/304/5679/1926.abstract}%
  \bibAnnoteFile{NoStop}{buszaki}%
\bibitem{cohen}%
  \BibitemOpen
  \bibfield{author}{%
  \bibinfo {author} {\bibfnamefont{A.~H.}\ \bibnamefont{Cohen}}, \bibinfo
  {author} {\bibfnamefont{P.~J.}\ \bibnamefont{Holmes}},\ and\ \bibinfo
  {author} {\bibfnamefont{R.~H.}\ \bibnamefont{Rand}},\ }%
  \bibfield{journal}{%
  \bibinfo {journal} {Journal of Mathematical Biology}\ }%
  \textbf{\bibinfo {volume} {13}},\ \bibinfo {pages} {345} (\bibinfo {year}
  {1982}),\ ISSN \bibinfo {issn} {0303-6812},\ \bibinfo {note}
  {10.1007/BF00276069},\ \url{http://dx.doi.org/10.1007/BF00276069}%
  \bibAnnoteFile{NoStop}{cohen}%
\bibitem{ermentrout_gelperin}%
  \BibitemOpen
  \bibfield{author}{%
  \bibinfo {author} {\bibfnamefont{B.}~\bibnamefont{Ermentrout}}, \bibinfo
  {author} {\bibfnamefont{J.}~\bibnamefont{Flores}},\ and\ \bibinfo {author}
  {\bibfnamefont{A.}~\bibnamefont{Gelperin}},\ }%
  \bibfield{journal}{%
  \bibinfo {journal} {Journal of Neurophysiology}\ }%
  \textbf{\bibinfo {volume} {79}},\ \bibinfo {pages} {2677} (\bibinfo {year}
  {1998}),\
  \Eprint{http://arxiv.org/abs/http://jn.physiology.org/content/79/5/2677.full%
.pdf+html}{http://jn.physiology.org/content/79/5/2677.full.pdf+html},\
  \url{http://jn.physiology.org/content/79/5/2677.abstract}%
  \bibAnnoteFile{NoStop}{ermentrout_gelperin}%
\bibitem{adaptation}%
  \BibitemOpen
  \bibfield{author}{%
  \bibinfo {author} {\bibfnamefont{B.}~\bibnamefont{Ermentrout}}, \bibinfo
  {author} {\bibfnamefont{M.}~\bibnamefont{Pascal}},\ and\ \bibinfo {author}
  {\bibfnamefont{B.}~\bibnamefont{Gutkin}},\ }%
  \bibfield{journal}{%
  \Doi{10.1162/08997660152002861}{\bibinfo {journal} {Neural Computation}}\ }%
  \textbf{\bibinfo {volume} {13}},\ \bibinfo {pages} {1285} (\bibinfo {year}
  {2001}),\
  \Eprint{http://arxiv.org/abs/http://www.mitpressjournals.org/doi/pdf/10.1162%
/08997660152002861}{http://www.mitpressjournals.org/doi/pdf/10.1162/0899766015%
2002861},\
  \url{http://www.mitpressjournals.org/doi/abs/10.1162/08997660152002861}%
  \bibAnnoteFile{NoStop}{adaptation}%
\bibitem{Mancilla}%
  \BibitemOpen
  \bibfield{author}{%
  \bibinfo {author} {\bibfnamefont{J.~G.}\ \bibnamefont{Mancilla}}, \bibinfo
  {author} {\bibfnamefont{T.~J.}\ \bibnamefont{Lewis}}, \bibinfo {author}
  {\bibfnamefont{D.~J.}\ \bibnamefont{Pinto}}, \bibinfo {author}
  {\bibfnamefont{J.}~\bibnamefont{Rinzel}},\ and\ \bibinfo {author}
  {\bibfnamefont{B.~W.}\ \bibnamefont{Connors}},\ }%
  \bibfield{journal}{%
  \Doi{10.1523/JNEUROSCI.2715-06.2007}{\bibinfo {journal} {The Journal of
  Neuroscience}}\ }%
  \textbf{\bibinfo {volume} {27}},\ \bibinfo {pages} {2058} (\bibinfo {year}
  {2007}),\
  \Eprint{http://arxiv.org/abs/http://www.jneurosci.org/content/27/8/2058.full%
.pdf+html}{http://www.jneurosci.org/content/27/8/2058.full.pdf+html},\
  \url{http://www.jneurosci.org/content/27/8/2058.abstract}%
  \bibAnnoteFile{NoStop}{Mancilla}%
\bibitem{Vreeswijk}%
  \BibitemOpen
  \bibfield{author}{%
  \bibinfo {author} {\bibfnamefont{C.}~\bibnamefont{Vreeswijk}}, \bibinfo
  {author} {\bibfnamefont{L.~F.}\ \bibnamefont{Abbott}},\ and\ \bibinfo
  {author} {\bibfnamefont{G.}~\bibnamefont{Bard~Ermentrout}},\ }%
  \bibfield{journal}{%
  \bibinfo {journal} {Journal of Computational Neuroscience}\ }%
  \textbf{\bibinfo {volume} {1}},\ \bibinfo {pages} {313} (\bibinfo {year}
  {1994}),\ ISSN \bibinfo {issn} {0929-5313},\ \bibinfo {note}
  {10.1007/BF00961879},\ \url{http://dx.doi.org/10.1007/BF00961879}%
  \bibAnnoteFile{NoStop}{Vreeswijk}%
\bibitem{Note1}%
  \BibitemOpen
  \bibinfo {note} {A silicon neuron refers to an electronic circuit designed to
  mimic a biological neuron}%
  \bibAnnoteFile{NoStop}{Note1}%
\bibitem{cymbalyuk}%
  \BibitemOpen
  \bibfield{author}{%
  \bibinfo {author} {\bibfnamefont{G.~S.}\ \bibnamefont{Cymbalyuk}}, \bibinfo
  {author} {\bibfnamefont{G.~N.}\ \bibnamefont{Patel}}, \bibinfo {author}
  {\bibfnamefont{R.~L.}\ \bibnamefont{Calabrese}}, \bibinfo {author}
  {\bibfnamefont{S.~P.}\ \bibnamefont{DeWeerth}},\ and\ \bibinfo {author}
  {\bibfnamefont{A.~H.}\ \bibnamefont{Cohen}},\ }%
  \bibfield{journal}{%
  \Doi{10.1162/089976600300014926}{\bibinfo {journal} {Neural Computation}}\ }%
  \textbf{\bibinfo {volume} {12}},\ \bibinfo {pages} {2259} (\bibinfo {year}
  {2000}),\
  \Eprint{http://arxiv.org/abs/http://www.mitpressjournals.org/doi/pdf/10.1162%
/089976600300014926}{http://www.mitpressjournals.org/doi/pdf/10.1162/089976600%
300014926},\
  \url{http://www.mitpressjournals.org/doi/abs/10.1162/089976600300014926}%
  \bibAnnoteFile{NoStop}{cymbalyuk}%
\bibitem{limax}%
  \BibitemOpen
  \bibfield{author}{%
  \bibinfo {author} {\bibfnamefont{A.}~\bibnamefont{Gelperin}},\ }%
  \bibfield{journal}{%
  \bibinfo {journal} {Journal of Experimental Biology}\ }%
  \textbf{\bibinfo {volume} {202}},\ \bibinfo {pages} {1855} (\bibinfo {year}
  {1999}),\
  \Eprint{http://arxiv.org/abs/http://jeb.biologists.org/content/202/14/1855.f%
ull.pdf+html}{http://jeb.biologists.org/content/202/14/1855.full.pdf+html},\
  \url{http://jeb.biologists.org/content/202/14/1855.abstract}%
  \bibAnnoteFile{NoStop}{limax}%
\bibitem{kim}%
  \BibitemOpen
  \bibfield{author}{%
  \bibinfo {author} {\bibfnamefont{U.}~\bibnamefont{Kim}}, \bibinfo {author}
  {\bibfnamefont{T.}~\bibnamefont{Bal}},\ and\ \bibinfo {author}
  {\bibfnamefont{D.~A.}\ \bibnamefont{McCormick}},\ }%
  \bibfield{journal}{%
  \bibinfo {journal} {Journal of Neurophysiology}\ }%
  \textbf{\bibinfo {volume} {74}},\ \bibinfo {pages} {1301} (\bibinfo {year}
  {1995}),\
  \Eprint{http://arxiv.org/abs/http://jn.physiology.org/content/74/3/1301.full%
.pdf+html}{http://jn.physiology.org/content/74/3/1301.full.pdf+html},\
  \url{http://jn.physiology.org/content/74/3/1301.abstract}%
  \bibAnnoteFile{NoStop}{kim}%
\bibitem{peinado}%
  \BibitemOpen
  \bibfield{author}{%
  \bibinfo {author} {\bibfnamefont{A.}~\bibnamefont{Peinado}},\ }%
  \bibfield{journal}{%
  \bibinfo {journal} {Journal of Neurophysiology}\ }%
  \textbf{\bibinfo {volume} {85}},\ \bibinfo {pages} {620} (\bibinfo {year}
  {2001}),\
  \Eprint{http://arxiv.org/abs/http://jn.physiology.org/content/85/2/620.full.%
pdf+html}{http://jn.physiology.org/content/85/2/620.full.pdf+html},\
  \url{http://jn.physiology.org/content/85/2/620.abstract}%
  \bibAnnoteFile{NoStop}{peinado}%
\bibitem{kleinfeld}%
  \BibitemOpen
  \bibfield{author}{%
  \bibinfo {author} {\bibfnamefont{G.}~\bibnamefont{Ermentrout}}\ and\ \bibinfo
  {author} {\bibfnamefont{D.}~\bibnamefont{Kleinfeld}},\ }%
  \bibfield{journal}{%
  \Doi{DOI: 10.1016/S0896-6273(01)00178-7}{\bibinfo {journal} {Neuron}}\ }%
  \textbf{\bibinfo {volume} {29}},\ \bibinfo {pages} {33 } (\bibinfo {year}
  {2001}),\ ISSN \bibinfo {issn} {0896-6273},\
  \url{http://www.sciencedirect.com/science/article/B6WSS-42D33KV-7/2/9571b90c%
c1aea3c7aa28c6ff54dea0c8}%
  \bibAnnoteFile{NoStop}{kleinfeld}%
\bibitem{ratspinalcordwaves}%
  \BibitemOpen
  \bibfield{author}{%
  \bibinfo {author} {\bibfnamefont{Y.}~\bibnamefont{Momose-Sato}}, \bibinfo
  {author} {\bibfnamefont{K.}~\bibnamefont{Sato}},\ and\ \bibinfo {author}
  {\bibfnamefont{M.}~\bibnamefont{Kinoshita}},\ }%
  \bibfield{journal}{%
  \Doi{10.1111/j.1460-9568.2007.05352.x}{\bibinfo {journal} {European Journal
  of Neuroscience}}\ }%
  \textbf{\bibinfo {volume} {25}},\ \bibinfo {pages} {929} (\bibinfo {year}
  {2007}),\ ISSN \bibinfo {issn} {1460-9568},\
  \url{http://dx.doi.org/10.1111/j.1460-9568.2007.05352.x}%
  \bibAnnoteFile{NoStop}{ratspinalcordwaves}%
\bibitem{bardbook}%
  \BibitemOpen
  \bibfield{author}{%
  \bibinfo {author} {\bibfnamefont{G.~B.}\ \bibnamefont{Ermentrout}}\ and\
  \bibinfo {author} {\bibfnamefont{D.}~\bibnamefont{Terman}},\ }%
  \emph{\bibinfo {title} {Mathematical Foundations of Neuroscience}}\ (\bibinfo
  {publisher} {Springer},\ \bibinfo {year} {2012})%
  \bibAnnoteFile{NoStop}{bardbook}%
\bibitem{chains}%
  \BibitemOpen
  \bibfield{author}{%
  \bibinfo {author} {\bibfnamefont{N.}~\bibnamefont{Kopell}}\ and\ \bibinfo
  {author} {\bibfnamefont{G.~B.}\ \bibnamefont{Ermentrout}},\ }%
  \bibfield{journal}{%
  \Doi{10.1002/cpa.3160390504}{\bibinfo {journal} {Communications on Pure and
  Applied Mathematics}}\ }%
  \textbf{\bibinfo {volume} {39}},\ \bibinfo {pages} {623} (\bibinfo {year}
  {1986}),\ ISSN \bibinfo {issn} {1097-0312},\
  \url{http://dx.doi.org/10.1002/cpa.3160390504}%
  \bibAnnoteFile{NoStop}{chains}%
\bibitem{Bressloff}%
  \BibitemOpen
  \bibfield{author}{%
  \bibinfo {author} {\bibfnamefont{P.~C.}\ \bibnamefont{Bressloff}}, \bibinfo
  {author} {\bibfnamefont{S.}~\bibnamefont{Coombes}},\ and\ \bibinfo {author}
  {\bibfnamefont{B.}~\bibnamefont{de~Souza}},\ }%
  \bibfield{journal}{%
  \Doi{10.1103/PhysRevLett.79.2791}{\bibinfo {journal} {Phys. Rev. Lett.}}\ }%
  \textbf{\bibinfo {volume} {79}},\ \bibinfo {pages} {2791} (\bibinfo {month}
  {Oct}\ \bibinfo {year} {1997}),\
  \url{http://link.aps.org/doi/10.1103/PhysRevLett.79.2791}%
  \bibAnnoteFile{NoStop}{Bressloff}%
\bibitem{jones2003}%
  \BibitemOpen
  \bibfield{author}{%
  \bibinfo {author} {\bibfnamefont{S.}~\bibnamefont{Jones}}, \bibinfo {author}
  {\bibfnamefont{B.}~\bibnamefont{Mulloney}}, \bibinfo {author}
  {\bibfnamefont{T.}~\bibnamefont{Kaper}},\ and\ \bibinfo {author}
  {\bibfnamefont{N.}~\bibnamefont{Kopell}},\ }%
  \bibfield{journal}{%
  \bibinfo {journal} {The Journal of neuroscience}\ }%
  \textbf{\bibinfo {volume} {23}},\ \bibinfo {pages} {3457} (\bibinfo {year}
  {2003})%
  \bibAnnoteFile{NoStop}{jones2003}%
\bibitem{wang}%
  \BibitemOpen
  \bibfield{author}{%
  \bibinfo {author} {\bibfnamefont{G.~B.}\ \bibnamefont{Xiao-Jing~Wang}},\ }%
  \bibfield{journal}{%
  \bibinfo {journal} {The Journal of Neuroscience}\ }%
  \textbf{\bibinfo {volume} {16}},\ \bibinfo {pages} {6402} (\bibinfo {month}
  {October}\ \bibinfo {year} {1996})%
  \bibAnnoteFile{NoStop}{wang}%
\bibitem{Note2}%
  \BibitemOpen
  \bibinfo {note} {\begin {math}\eta =Q_{10}^{T-T_{base}}\end {math}: in this
  equations, $Q_{0}$ is the ``ratio of the rates for an increase in temperature
  of $10^0$C'' \cite {bardbook}.}%
  \bibAnnoteFile{Stop}{Note2}%
\bibitem{Pfeuty}%
  \BibitemOpen
  \bibfield{author}{%
  \bibinfo {author} {\bibfnamefont{B.}~\bibnamefont{Pfeuty}}, \bibinfo {author}
  {\bibfnamefont{G.}~\bibnamefont{Mato}}, \bibinfo {author}
  {\bibfnamefont{D.}~\bibnamefont{Golomb}},\ and\ \bibinfo {author}
  {\bibfnamefont{D.}~\bibnamefont{Hansel}},\ }%
  \bibfield{journal}{%
  \bibinfo {journal} {The Journal of Neuroscience}\ }%
  \textbf{\bibinfo {volume} {23}},\ \bibinfo {pages} {6280} (\bibinfo {year}
  {2003}),\
  \Eprint{http://arxiv.org/abs/http://www.jneurosci.org/content/23/15/6280.ful%
l.pdf+html}{http://www.jneurosci.org/content/23/15/6280.full.pdf+html},\
  \url{http://www.jneurosci.org/content/23/15/6280.abstract}%
  \bibAnnoteFile{NoStop}{Pfeuty}%
\bibitem{Nusbaum}%
  \BibitemOpen
  \bibfield{author}{%
  \bibinfo {author} {\bibfnamefont{M.~P.}\ \bibnamefont{{Nusbaum}}}\ and\
  \bibinfo {author} {\bibfnamefont{M.~P.}\ \bibnamefont{{Beenhakker}}},\ }%
  \bibfield{journal}{%
  \bibinfo {journal} {\nat}\ }%
  \textbf{\bibinfo {volume} {417}},\ \bibinfo {pages} {343} (\bibinfo {month}
  {May}\ \bibinfo {year} {2002})%
  \bibAnnoteFile{NoStop}{Nusbaum}%
\bibitem{javedan}%
  \BibitemOpen
  \bibfield{author}{%
  \bibinfo {author} {\bibfnamefont{S.~P.}\ \bibnamefont{Javedan}}, \bibinfo
  {author} {\bibfnamefont{R.~S.}\ \bibnamefont{Fisher}}, \bibinfo {author}
  {\bibfnamefont{H.~G.}\ \bibnamefont{Eder}}, \bibinfo {author}
  {\bibfnamefont{K.}~\bibnamefont{Smith}},\ and\ \bibinfo {author}
  {\bibfnamefont{J.}~\bibnamefont{Wu}},\ }%
  \bibfield{journal}{%
  \bibinfo {journal} {Epilepsia}\ }%
  \textbf{\bibinfo {volume} {46}},\ \bibinfo {pages} {574} (\bibinfo {year}
  {2002})%
  \bibAnnoteFile{NoStop}{javedan}%
\bibitem{strogatz}%
  \BibitemOpen
  \bibfield{author}{%
  \bibinfo {author} {\bibfnamefont{S.}~\bibnamefont{H.}}\ and\ \bibinfo
  {author} {\bibnamefont{Strogatz}},\ }%
  \bibfield{journal}{%
  \Doi{10.1016/S0167-2789(00)00094-4}{\bibinfo {journal} {Physica D: Nonlinear
  Phenomena}}\ }%
  \textbf{\bibinfo {volume} {143}},\ \bibinfo {pages} {1 } (\bibinfo {year}
  {2000}),\ ISSN \bibinfo {issn} {0167-2789},\
  \url{http://www.sciencedirect.com/science/article/pii/S0167278900000944}%
  \bibAnnoteFile{NoStop}{strogatz}%
\bibitem{kelso}%
  \BibitemOpen
  \bibfield{author}{%
  \bibinfo {author} {\bibfnamefont{H.}~\bibnamefont{Haken}}, \bibinfo {author}
  {\bibfnamefont{J.}~\bibnamefont{Kelso}},\ and\ \bibinfo {author}
  {\bibfnamefont{H.}~\bibnamefont{Bunz}},\ }%
  \bibfield{journal}{%
  \bibinfo {journal} {Biological cybernetics}\ }%
  \textbf{\bibinfo {volume} {51}},\ \bibinfo {pages} {347} (\bibinfo {year}
  {1985})%
  \bibAnnoteFile{NoStop}{kelso}%
\bibitem{ipp}%
  \BibitemOpen
  \bibfield{author}{%
  \bibinfo {author} {\bibfnamefont{G.~B.}\ \bibnamefont{Ermentrout}}\ and\
  \bibinfo {author} {\bibfnamefont{N.}~\bibnamefont{Kopell}},\ }%
  \bibfield{journal}{%
  \bibinfo {journal} {SIAM Journal on Applied Mathematics}\ }%
  \textbf{\bibinfo {volume} {54}},\ \bibinfo {pages} {pp. 478} (\bibinfo {year}
  {1994}),\ ISSN \bibinfo {issn} {00361399},\
  \url{https://sremote.pitt.edu:11017/stable/2102230}%
  \bibAnnoteFile{NoStop}{ipp}%
\bibitem{sync}%
  \BibitemOpen
  \bibfield{author}{%
  \bibinfo {author} {\bibfnamefont{D.~A.}\ \bibnamefont{Wiley}}, \bibinfo
  {author} {\bibfnamefont{S.~H.}\ \bibnamefont{Strogatz}},\ and\ \bibinfo
  {author} {\bibfnamefont{M.}~\bibnamefont{Girvan}},\ }%
  \bibfield{journal}{%
  \Doi{DOI:10.1063/1.2165594}{\bibinfo {journal} {CHAOS}}\ }%
  \textbf{\bibinfo {volume} {16}},\ \bibinfo {pages} {015103} (\bibinfo {year}
  {2006}),\ ISSN \bibinfo {issn} {10541500},\
  \url{http://dx.doi.org/10.1063/1.2165594}%
  \bibAnnoteFile{NoStop}{sync}%
\bibitem{blasius}%
  \BibitemOpen
  \bibfield{author}{%
  \bibinfo {author} {\bibfnamefont{B.}~\bibnamefont{Blasius}}\ and\ \bibinfo
  {author} {\bibfnamefont{R.}~\bibnamefont{T\"onjes}},\ }%
  \bibfield{journal}{%
  \Doi{10.1103/PhysRevLett.95.084101}{\bibinfo {journal} {Phys. Rev. Lett.}}\
  }%
  \textbf{\bibinfo {volume} {95}},\ \bibinfo {pages} {084101} (\bibinfo {month}
  {Aug}\ \bibinfo {year} {2005}),\
  \url{http://link.aps.org/doi/10.1103/PhysRevLett.95.084101}%
  \bibAnnoteFile{NoStop}{blasius}%
\bibitem{Note3}%
  \BibitemOpen
  \bibinfo {note} {These are not to be confused with fractured waves studied by
  Kopell (also known as s-waves) \cite {somers}.}%
  \bibAnnoteFile{Stop}{Note3}%
\bibitem{PhysRevLett.68.1730}%
  \BibitemOpen
  \bibfield{author}{%
  \bibinfo {author} {\bibfnamefont{B.}~\bibnamefont{Denardo}}, \bibinfo
  {author} {\bibfnamefont{B.}~\bibnamefont{Galvin}}, \bibinfo {author}
  {\bibfnamefont{A.}~\bibnamefont{Greenfield}}, \bibinfo {author}
  {\bibfnamefont{A.}~\bibnamefont{Larraza}}, \bibinfo {author}
  {\bibfnamefont{S.}~\bibnamefont{Putterman}},\ and\ \bibinfo {author}
  {\bibfnamefont{W.}~\bibnamefont{Wright}},\ }%
  \bibfield{journal}{%
  \Doi{10.1103/PhysRevLett.68.1730}{\bibinfo {journal} {Phys. Rev. Lett.}}\ }%
  \textbf{\bibinfo {volume} {68}},\ \bibinfo {pages} {1730} (\bibinfo {month}
  {Mar}\ \bibinfo {year} {1992}),\
  \url{http://link.aps.org/doi/10.1103/PhysRevLett.68.1730}%
  \bibAnnoteFile{NoStop}{PhysRevLett.68.1730}%
\bibitem{dogfish}%
  \BibitemOpen
  \bibfield{author}{%
  \bibinfo {author} {\bibfnamefont{S.}~\bibnamefont{Grillner}},\ }%
  \bibfield{journal}{%
  \bibinfo {journal} {Experimental Brain Research}\ }%
  \textbf{\bibinfo {volume} {20}},\ \bibinfo {pages} {459} (\bibinfo {year}
  {1974})%
  \bibAnnoteFile{NoStop}{dogfish}%
\bibitem{cat}%
  \BibitemOpen
  \bibfield{author}{%
  \bibinfo {author} {\bibfnamefont{J.}~\bibnamefont{Christensen}}\ and\
  \bibinfo {author} {\bibfnamefont{R.}~\bibnamefont{Hauser}},\ }%
  \bibfield{journal}{%
  \bibinfo {journal} {American Journal of Physiology -- Legacy Content}\ }%
  \textbf{\bibinfo {volume} {221}},\ \bibinfo {pages} {1033} (\bibinfo {year}
  {1971}),\
  \Eprint{http://arxiv.org/abs/http://ajplegacy.physiology.org/content/221/4/1%
033.full.pdf+html}{http://ajplegacy.physiology.org/content/221/4/1033.full.pdf%
+html},\ \url{http://ajplegacy.physiology.org/content/221/4/1033.short}%
  \bibAnnoteFile{NoStop}{cat}%
\bibitem{KevinW}%
  \BibitemOpen
  \bibfield{author}{%
  \bibinfo {author} {\bibfnamefont{K.~W.}\ \bibnamefont{Thompson}},\ }%
  \bibfield{journal}{%
  \Doi{10.1016/0021-9991(87)90041-6}{\bibinfo {journal} {Journal of
  Computational Physics}}\ }%
  \textbf{\bibinfo {volume} {68}},\ \bibinfo {pages} {1 } (\bibinfo {year}
  {1987}),\ ISSN \bibinfo {issn} {0021-9991},\
  \url{http://www.sciencedirect.com/science/article/pii/0021999187900416}%
  \bibAnnoteFile{NoStop}{KevinW}%
\bibitem{Note4}%
  \BibitemOpen
  \bibinfo {note} {To see that this is consistent with the preceding statement,
  start with the $H(-\phi _{j-1})=H(\phi _{j})$, using a Taylor series to
  expand out each side and evaluating at the wave solution, we have the
  expression: \begin {eqnarray} H'(-\phi )\protect \frac {d\phi }{dx}=0 \end
  {eqnarray} Since $H'(\phi )\not =0$, then it must be that $(\protect \frac
  {d\phi }{dx}|_{x=0})=0$.}%
  \bibAnnoteFile{Stop}{Note4}%
\bibitem{compacton1}%
  \BibitemOpen
  \bibfield{author}{%
  \bibinfo {author} {\bibfnamefont{A.}~\bibnamefont{Pikovsky}}\ and\ \bibinfo
  {author} {\bibfnamefont{P.}~\bibnamefont{Rosenau}},\ }%
  \bibfield{journal}{%
  \Doi{10.1016/j.physd.2006.04.015}{\bibinfo {journal} {Physica D: Nonlinear
  Phenomena}}\ }%
  \textbf{\bibinfo {volume} {218}},\ \bibinfo {pages} {56 } (\bibinfo {year}
  {2006}),\ ISSN \bibinfo {issn} {0167-2789},\
  \url{http://www.sciencedirect.com/science/article/pii/S0167278906001382}%
  \bibAnnoteFile{NoStop}{compacton1}%
\bibitem{compacton2}%
  \BibitemOpen
  \bibfield{author}{%
  \bibinfo {author} {\bibfnamefont{K.}~\bibnamefont{Ahnert}}\ and\ \bibinfo
  {author} {\bibfnamefont{A.}~\bibnamefont{Pikovsky}},\ }%
  \bibfield{journal}{%
  \Doi{10.1103/PhysRevE.79.026209}{\bibinfo {journal} {Phys. Rev. E}}\ }%
  \textbf{\bibinfo {volume} {79}},\ \bibinfo {pages} {026209} (\bibinfo {month}
  {Feb}\ \bibinfo {year} {2009}),\
  \url{http://link.aps.org/doi/10.1103/PhysRevE.79.026209}%
  \bibAnnoteFile{NoStop}{compacton2}%
\bibitem{compacton3}%
  \BibitemOpen
  \bibfield{author}{%
  \bibinfo {author} {\bibfnamefont{K.}~\bibnamefont{{Ahnert}}}\ and\ \bibinfo
  {author} {\bibfnamefont{A.}~\bibnamefont{{Pikovsky}}},\ }%
  \bibfield{journal}{%
  \Doi{10.1063/1.2955758}{\bibinfo {journal} {Chaos}}\ }%
  \textbf{\bibinfo {volume} {18}},\ \bibinfo {pages} {037118} (\bibinfo {month}
  {Sep.}\ \bibinfo {year} {2008}),\
  \Eprint{http://arxiv.org/abs/0806.1833}{arXiv:0806.1833 [nlin.PS]}%
  \bibAnnoteFile{NoStop}{compacton3}%
\bibitem{Alex}%
  \BibitemOpen
  \bibfield{author}{%
  \bibinfo {author} {\bibfnamefont{A.}~\bibnamefont{Urban}},\ }%
  \emph{\bibinfo {title} {Intermediate Stable Phase Locked States in Oscillator
  Networks}},\ Ph.D. thesis,\ \bibinfo {school} {University of Pittsburgh}
  (\bibinfo {year} {2011})%
  \bibAnnoteFile{NoStop}{Alex}%
\bibitem{Note5}%
  \BibitemOpen
  \bibinfo {note} {Gershgorin Circle Theorem: Let:$ A=[a_{ij}]$ be an arbitrary
  n x n matrix with elements that may be complex and let:\\ \begin {eqnarray}
  \Lambda _{i}=\DOTSB \sum@ \slimits@ ^{n}_{j=1,i\not =j}|a_{ij}|\protect
  \hspace {.25in} for\protect \hspace {.1in} i=1,2,...n \end {eqnarray} Then
  all of the eigenvalues $\lambda _{i}$ of$A$ lie in the union of n disks
  $\Gamma _{i}$ where: \begin {eqnarray} \Gamma _{i}:|\lambda -a_{ii}|\leq
  |\lambda _{i}\protect \hspace {.25in} for\protect \hspace {.1in} i=1,2,...n
  \end {eqnarray} This wording of the Gershgorin Circle theorem was taken
  from:\\ Tables of Integrals, Series, and Products by I.S.Gradshteyn and I.M.
  Ryzhik\cite {integrals}.}%
  \bibAnnoteFile{Stop}{Note5}%
\bibitem{Note6}%
  \BibitemOpen
  \bibinfo {note} {All equations in this section were integrated with Euler's
  method. The step size used is $.01$.}%
  \bibAnnoteFile{Stop}{Note6}%
\bibitem{Note7}%
  \BibitemOpen
  \bibinfo {note} {The Fourier terms used may or may not be calculated from the
  full model. For instance, an interaction function with an artificially
  inflated second odd mode will produce an interesting fractured pattern,
  whereas many interaction functions computed from the full model will not.}%
  \bibAnnoteFile{Stop}{Note7}%
\bibitem{somers}%
  \BibitemOpen
  \bibfield{author}{%
  \bibinfo {author} {\bibfnamefont{D.}~\bibnamefont{Somers}}\ and\ \bibinfo
  {author} {\bibfnamefont{N.}~\bibnamefont{Kopell}},\ }%
  \bibfield{journal}{%
  \Doi{10.1016/0167-2789(95)00198-0}{\bibinfo {journal} {Physica D: Nonlinear
  Phenomena}}\ }%
  \textbf{\bibinfo {volume} {89}},\ \bibinfo {pages} {169 } (\bibinfo {year}
  {1995}),\ ISSN \bibinfo {issn} {0167-2789},\
  \url{http://www.sciencedirect.com/science/article/pii/0167278995001980}%
  \bibAnnoteFile{NoStop}{somers}%
\bibitem{integrals}%
  \BibitemOpen
  \bibfield{author}{%
  \bibinfo {author} {\bibfnamefont{I.}~\bibnamefont{Gradshteyn}}\ and\ \bibinfo
  {author} {\bibfnamefont{I.}~\bibnamefont{Ryzhik}},\ }%
  \emph{\bibinfo {title} {Tables of Integrals, Series and Products}}\ (\bibinfo
  {publisher} {Academic Press},\ \bibinfo {year} {2000})%
  \bibAnnoteFile{NoStop}{integrals}%
\end{thebibliography}%


%
\bibliographystyle{apsrev4-1}
\end{document}